\documentclass[aps,pre,reprint,superscriptaddress,longbibliography]{revtex4-2}
\usepackage{tikz}
\usepackage[percent]{overpic}
\usepackage[T1]{fontenc}
\usepackage[utf8]{inputenc}
\usepackage{lmodern}
\usepackage{microtype}
\usepackage{amsmath,amssymb}
\usepackage{graphicx}
\usepackage[caption=false]{subfig}
\usepackage{hyperref}
\hypersetup{hidelinks}
\usepackage{xcolor}
\usepackage{comment}
\usepackage{siunitx}

\newcommand{\xizt}{\Xi(z,T)}
\newcommand{\znt}{Z_N(T)}

\newcommand{\fluct}{\operatorname{Var}}
\newcommand{\meanv}[1]{\langle {#1} \rangle}
\newcommand{\avgmce}[1]{\left\langle {#1} \right\rangle_{N,E}}
\newcommand{\avgce}[1]{\left\langle {#1} \right\rangle_N}
\newcommand{\avggce}[1]{\left\langle {#1} \right\rangle}

\begin{document}
\title{Ensemble-dependent condensate statistics and correlations in an ideal Bose gas}

\author{S. Augustynowicz}
\affiliation{Center for Theoretical Physics, Polish Academy of Sciences,
Al. Lotników 32/46, 02-668 Warsaw, Poland}

\author{J. Okoński}
\affiliation{Faculty of Physics, University of Warsaw,
Pasteura 5, 02-093 Warsaw, Poland}
\affiliation{Center for Theoretical Physics, Polish Academy of Sciences,
Al. Lotników 32/46, 02-668 Warsaw, Poland}

\author{M. B. Kruk}
\affiliation{Institute of Physics, Polish Academy of Sciences,
Aleja Lotników 32/46, 02-668 Warsaw, Poland}
\affiliation{Center for Theoretical Physics, Polish Academy of Sciences,
Al. Lotników 32/46, 02-668 Warsaw, Poland}

\author{I. Zebergs}
\affiliation{Center for Complex Quantum Systems,
Department of Physics and Astronomy, Aarhus University,
Ny Munkegade 120, DK-8000 Aarhus C, Denmark}

\author{J. J. Arlt}
\affiliation{Center for Complex Quantum Systems, Department of Physics and Astronomy, Aarhus University, Ny Munkegade 120, DK-8000 Aarhus C, Denmark}

\author{K. Rzążewski}
\affiliation{Center for Theoretical Physics, Polish Academy of Sciences,
Al. Lotników 32/46, 02-668 Warsaw, Poland}
\author{K. Pawłowski}
\email{pawlowski@cft.edu.pl}
\affiliation{Center for Theoretical Physics, Polish Academy of Sciences,
Al. Lotników 32/46, 02-668 Warsaw, Poland}

\date{\today}

\begin{abstract}
The choice of statistical ensemble has a measurable impact on the statistical description of Bose--Einstein condensates beyond the well-known grand canonical anomaly. We develop a segmented-contour approach for the ideal Bose gas, which refines the saddle-point approximation by including non-Gaussian corrections in the canonical and microcanonical ensembles. The method applies to a broad class of noninteracting spectra, and provides efficient access to condensate fluctuations, higher moments, and correlation functions.
For a three-dimensional harmonic trap with fixed anisotropy, we find that the ratio of the maxima of the microcanonical and canonical condensate-number variances approaches the theoretically predicted asymptotic value $1-3\zeta(3)^2/[4\zeta(2)\zeta(4)]\approx 0.391$
only extremely slowly with increasing particle number. We also show that ensemble effects are visible in the second-order correlation function $g^{(2)}(x)$ near the condensation crossover. After ballistic release from a three-dimensional harmonic trap, the time evolution of $g^{(2)}(x)$ preserves the differences between statistical ensembles.
\end{abstract}

\maketitle

\section{Introduction}
Physical systems at finite temperature are typically described using three statistical ensembles: the microcanonical (MCE), canonical (CE), and grand canonical (GCE) ensembles. While these formalisms represent systems with different degrees of isolation from their environment, they are frequently used interchangeably in computational practice. This practice is supported by classic textbook arguments for ensemble equivalence~\cite{Kubo1965StatMech,Pathria2011,Huang1987,
Landau1980,Touchette15}, which, however, apply primarily to standard extensive and intensive quantities. In contrast, higher-order quantities such as condensate particle-number fluctuations or correlation functions do not fall under these equivalence theorems~\cite{ziff1977,Kruk2025Fluctuations} and may differ substantially between ensembles.

The inequivalence of statistical ensembles in the context of Bose--Einstein condensation has been the subject of numerous theoretical studies~\cite{ziff1977,Politzer1996,
WeissWilkens1997Counting,Navez1997FourthEnsemble,
Holthaus1998,Kocharovsky2000}, summarized in review articles published in 2006~\cite{KocharovskyReview2006} and 2025~\cite{Kruk2025Fluctuations}. Historically, these efforts focused on the fluctuations of the number of condensed atoms, mostly in the ideal Bose gas. For a finite ideal Bose gas in a box, Glaum, Kleinert,
and Pelster derived canonical recurrence relations
and analyzed finite-size effects~\cite{GlaumKleinertPelster2007}. Even for a noninteracting gas, calculating partition functions for a fully isolated system, i.e., in the microcanonical ensemble, is demanding \cite{Grossmann1996,Gajda97,
Navez1997FourthEnsemble,Holthaus1998,Kruk23}
and closely related to classical problems in number theory, such as integer partitions~\cite{HardyRamanujan1918,Holthaus1998,Kruk2025Fluctuations}.  More recently, Fock-state sampling has provided
an alternative numerical approach to canonical
and microcanonical condensate fluctuations,
including applications to interacting
Bose gases~\cite{Kruk23,KrukVibel2023FSS}. Recent theoretical studies have also explored
the relationship between statistical ensembles,
including a continuous interpolation between
canonical and grand canonical classical-field
descriptions~\cite{Pietraszewicz2017}.
The simplest setting -- an ideal Bose gas in the grand canonical ensemble -- has long puzzled physicists because of its anomalously large condensate-number fluctuations, known as the grand canonical catastrophe ~\cite{ziff1977,KocharovskyReview2006,
Zannetti2015,Crisanti2019}.
The physical interpretation of these fluctuations,
particularly in connection with spontaneous
symmetry breaking, remains a subject of discussion
~\cite{Yukalov24,Crisanti2026}.

Interest in the ensemble-dependent quantities has revived recently %
due to the first direct experimental observations of atomic condensate particle-number fluctuations~\cite{Kristensen2019}, the subsequent demonstration of their microcanonical nature~\cite{Arlt2021}, the recent first measurement of BEC fluctuations in an optical lattice~\cite{JalaliMola2026} and previous experimental results on the photonic BEC ~\cite{klaersBoseEinsteinCondensation2010a},
where grand-canonical condensate-number statistics
have been directly observed~\cite{Schmitt14}. The observed microcanonical nature is well justified from a physical standpoint: ultracold atomic gases are produced under ultra-high vacuum conditions, nearly perfectly isolated from their environment. They therefore constitute textbook realizations of closed quantum systems, with residual fluctuations of energy and particle number inherited from the cooling process rather than introduced by coupling to an external reservoir. Notably, Refs.~\cite{Kristensen2019,Arlt2021} studied weakly interacting gases near the critical temperature, where the condensate fraction was small and the ideal-gas description provided a useful reference.

Understanding the differences arising from the choice of statistical ensemble has become even more important due to recent technological breakthroughs. Bose--Einstein condensates now play a central role in matter-wave interferometry~\cite{Cronin2009}, including high-precision gravimetry~\cite{Peters1999,Peters2001}.
At the same time, rapid advances in detection  techniques, most notably matter-wave microscopy, enable exceptionally precise measurements of physical observables, particularly correlation functions~\cite{deJongh2025,Yao2025,Xiang2025,Paquiez2026}.

Correlations in interacting one-dimensional Bose gases
have been studied extensively. In particular, local
pair correlations and their dependence on temperature
and interaction strength have been investigated
within the Lieb--Liniger model~\cite{Kheruntsyan03},
while nonlocal pair correlations and dynamical
density response functions have been examined
at finite temperature~\cite{Deuar09,Panfil2014}.
Canonical condensate statistics and low-order
correlation functions have also been investigated
in trapped one-dimensional gases~\cite{Bienias11a}.

Nevertheless, the dependence of spatial correlation functions on the choice of statistical ensemble, particularly the distinction between the canonical and microcanonical ensembles, remains much less explored. This question is especially relevant for density--density correlations, which are sensitive to condensate-number fluctuations. In fact, we show that  different ensembles yield even strikingly different physical pictures at low temperatures: strong bunching in the grand canonical ensemble, and nearly Poissonian or sub-Poissonian, laser-like fluctuations in the canonical and microcanonical ensembles. A complete analysis should include the experimentally accessible correlation functions: $g^{(1)}$, which characterizes phase coherence and can be probed interferometrically; and $g^{(2)}$ and $g^{(3)}$, which govern two-body and three-body loss rates, across different temperature regimes. 
\begin{figure}
\includegraphics[width=0.48\textwidth]{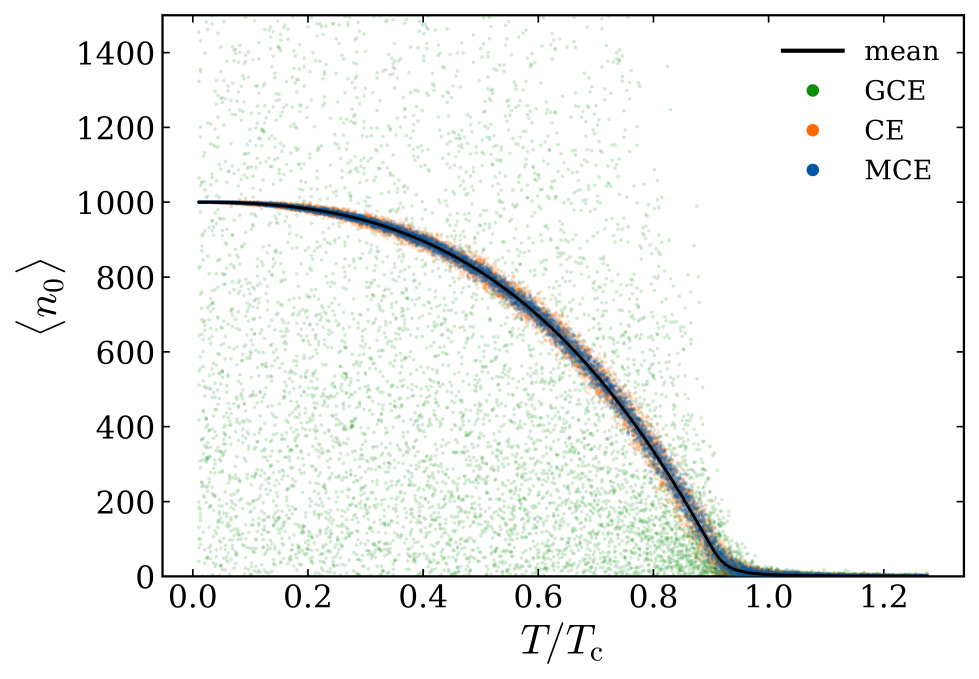}
\caption{Ensemble dependence of condensate-number fluctuations for an ideal Bose gas with $N=1000$ in the canonical and microcanonical ensembles and $\langle N\rangle=1000$ in the grand canonical ensemble. Dots are simulated single realizations of the condensate occupation $n_0$ (20 shots per temperature) in the grand canonical (green),
canonical (orange), and microcanonical (blue) ensembles. The mean condensate occupation is the same to leading order in \(N\) in all three ensembles and is shown as a single black line. The spread of the individual realizations is not: the grand canonical ensemble scatters over a range of order $N$ even well below $T_{\rm c}$, whereas fixing the atom number, and then also the energy, suppresses the scatter successively. Temperature is measured in units of $ T_{\rm c}=(\hbar\omega/k_{\rm B})
[N/\zeta(3)]^{1/3}$.}
\label{fig:graphical-abstract}
\end{figure}
This makes $g^{(2)}$ a good choice for an observable suitable for studying the inequivalence of statistical ensembles, since the ensembles may give almost identical mean occupations but different higher moments of $n_0$. This is illustrated in Fig.~\ref{fig:graphical-abstract}, where single realizations of $n_0$ in the three ensembles are shown to share the same mean but differ strikingly in their spread. %

In this work, we focus on the statistics of condensate occupation $n_0$ and the corresponding correlation function $g^{(2)}$ for the ideal Bose gas in different statistical ensembles. Along the way, we develop an efficient segmented-contour method for the canonical and microcanonical ensembles and benchmark it against previous approaches and exact recurrence results where available. The method remains efficient and accurate for large systems, including cases in which the condensate statistics are non-Gaussian. The resulting ensemble dependence in the ideal gas is purely statistical; in real systems, it will be supplemented by interaction- and dynamics-induced effects.

The paper is organized as follows. Section~\ref{sec:System} defines the system, the three statistical ensembles and the correlation functions. Section~\ref{sec:saddle-limitations} recalls the saddle-point method and shows where it fails, and Sec.~\ref{sec:gaussianity} introduces the cumulants as a convenient measure of the Gaussianity of distributions. The segmented-contour construction is developed for the
canonical ensemble in Sec.~\ref{subsec:segmented-ZN} and extended to the microcanonical one in Sec.~\ref{subsec:microocanonical}. We then discuss the Gaussian limit, the ratio of maximal fluctuations in the two ensembles, the ensemble dependence of $g^{(2)}$, and its behaviour after ballistic expansion. Technical details of the segmented-contour construction are discussed in Appendix~\ref{app:segmented-method}.

\section{Physical system, statistical ensembles, and observables of interest}
\label{sec:System}
We consider a noninteracting gas of bosons in $d$ spatial dimensions localized in a given region of space and characterized by a single-particle spectrum $\{\varepsilon_j\}$. Without loss of generality, we assume the lowest energy state to have energy equal to zero, $\varepsilon_0 = 0$. The many-body eigenstates are specified by the occupation numbers $\{n_j\}$ of the single-particle states, with energies and total number of atoms given by
\begin{align}
E\left(\{n_j\}\right) = \sum_j n_j \varepsilon_j,
\quad
N\left(\{n_j\}\right) = \sum_j n_j ,
\label{eq:E-N_constraints}
\end{align}
where  $n_j = 0,1,2,\dots$. 
We do not restrict the single-particle spectrum to any particular form and, in particular, we do not require the single-particle spectrum to be commensurate, i.e., the energies need not be expressible as $\varepsilon_j=m_j\delta\varepsilon$ for some common spacing $\delta\varepsilon>0$ and integers $m_j\geq0$. However, the numerical benchmarks in this work focus on commensurate
spectra, for which independent recurrence calculations are available. We focus on systems at thermal equilibrium within the three statistical ensembles introduced
below.

Conceptually, the simplest statistical ensemble is the microcanonical
statistical ensemble (MCE), describing an isolated system with exchange
of neither energy nor particles with its surroundings. Following the
standard construction of Kubo \cite{Kubo1965StatMech}, for a finite
system we specify the energy within a narrow interval
$[E,E+\delta E)$ and fix the total number of particles to $N$.
The thermodynamic weight $W_N(E,\delta E)$ is the number of many-body
states satisfying these constraints. According to the principle of
equal a priori probabilities, all such states are equally probable, and
the corresponding entropy is
\begin{equation}
S_N(E,\delta E)
=
k_{\rm B}\ln W_N(E,\delta E).
\label{eq:MCE-entropy}
\end{equation}

It is convenient to label each many-body configuration by
$\alpha = \{n_j^{(\alpha)}\}$, with
$E_\alpha = E(\{n_j^{(\alpha)}\})$ and
$N_\alpha =  N(\{n_j^{(\alpha)}\})$.
The cumulative number of many-body states with fixed particle number
$N$ is then
\begin{equation}
\mathcal Q_N(E)
=
\sum_{\alpha:\,N_\alpha=N}
\Theta\!\left(E-E_\alpha\right)
\label{eq:QN-cumulative}
\end{equation}
where $\Theta$ denotes the Heaviside step function. The thermodynamic weight of the energy shell can therefore be written
as
\begin{equation}
W_N(E,\delta E)
=
\mathcal Q_N(E+\delta E)-\mathcal Q_N(E).
\label{eq:W-Q}
\end{equation}
We consider an energy shell $[E,E+\delta E)$ containing many many-body levels while remaining narrow on the macroscopic energy scale. Averaging the exact spectral density over such a shell defines its smooth thermodynamic envelope, for which
\begin{equation}
W_N(E,\delta E)
\simeq
\Omega_N(E)\,\delta E .
\label{eq:W-Omega}
\end{equation}
Throughout this work, we use the same symbol $\Omega_N(E)$ for both the exact density of states, understood in the distributional sense when obtained by inverse Laplace transformation, and its smooth coarse-grained thermodynamic envelope when level-scale structure is irrelevant. Accordingly, equations involving inverse Laplace transforms are exact distributional identities, whereas the contour approximation evaluates the corresponding smooth envelope. As emphasized by Kubo~\cite{Kubo1965StatMech}, the particular choice of $\delta E$ does not affect the thermodynamic properties of a macroscopic system, as long as the energy shell contains many states while remaining narrow on the macroscopic energy scale. 

As recently demonstrated~\cite{Arlt2021}, the microcanonical ensemble provides a faithful description of typical cold-atom experiments. This results from the fact that atoms are effectively isolated from their surroundings. We also consider the canonical and grand canonical ensembles, which are computationally simpler and are the starting point for our microcanonical construction. The canonical ensemble describes a system in thermal equilibrium with a heat reservoir at temperature $T$. In this case, the configurations $\{n_j\}$ are constrained to total particle number $N$, while the total energy $E$ is allowed to vary. The canonical partition function is the Laplace transform of the fixed-$N$ state density,
\begin{equation}
Z_N(\beta)
=
\int_0^\infty \Omega_N(E) \, e^{-\beta E}\,dE .
\label{eq:Zn-Omega}
\end{equation}

Finally, in the grand canonical ensemble, the subsystem can exchange both energy and particles with a reservoir at temperature $T$ and chemical potential $\mu$. The relation between the canonical partition function $\znt$ and the
grand canonical one $\xizt$ is
\begin{equation}
\xizt = \sum_{N = 0}^{\infty} z^{N}\,\znt, 
\label{eq:Xi-Zn}
\end{equation}
where $z=e^{\beta\mu}$ is the fugacity.
In contrast to the other ensembles, the partition function $\xizt$ admits a simple explicit form for an ideal Bose gas with an arbitrary single-particle spectrum
\begin{equation}
\xizt
=
\prod_j
\sum_{n_j=0}^{\infty}
\exp\!\left[-\beta(\varepsilon_j-\mu)n_j\right]
=
\prod_j
\frac{1}{1-z e^{-\beta\varepsilon_j}},
\label{eqn:xizt}
\end{equation}
where the index $j$ runs over all single-particle states, including degeneracies. Each factor in Eq.~\eqref{eqn:xizt} has a pole at $z=e^{\beta\varepsilon_j}$; in particular, the ground-state contribution has a pole at $z=1$.

For finite-size systems, exact recurrence relations provide useful numerical benchmarks for the canonical and microcanonical ensembles. In the canonical ensemble, the partition function satisfies the Wilkens-Weiss recurrence relation
~\cite{WeissWilkens1997Counting,Landsberg1961,BorrmannFranke1993}
\begin{equation}
Z_N(\beta)
=
\frac{1}{N}
\sum_{\ell=1}^{N}
Z_1(\ell\beta)\,
Z_{N-\ell}(\beta),
\qquad
Z_0=1.
\label{eq:canonical-recurrence}
\end{equation}
For spectra lying on a discrete energy grid with spacing $\delta\varepsilon$, an exact recurrence is also available in the microcanonical case. Choosing the energy shell as
$[E,E+\delta\varepsilon)$, with $E$ restricted to the grid, the
thermodynamic weight $W_N(E,\delta\varepsilon)$ obeys
\begin{equation}
W_N(E,\delta\varepsilon)
=
\frac{1}{N}
\sum_{\ell=1}^{N}
\sum_j
W_{N-\ell}
\!\left(E-\ell\varepsilon_j,\delta\varepsilon\right),
\label{eq:microcanonical-recurrence}
\end{equation}
with $W_0(E,\delta\varepsilon)=\delta_{E,0}$ and
$W_N(E,\delta\varepsilon)=0$ for energies outside the $N$-particle spectrum
\cite{WeissWilkens1997Counting,ChaseMekjianZamick1999}.

We introduce a notation for expectation values in which the subscripts indicate the quantities held fixed by the ensemble: no subscript for the grand canonical ensemble, $N$ for the canonical ensemble, and $N,E$ for the microcanonical ensemble. Accordingly, expectation values are denoted by $\avggce{\,\cdot\,}$, $\avgce{\,\cdot\,}$, and $\avgmce{\,\cdot\,}$, respectively. In the GCE and CE, the mean ground-state
occupations are given by
\begin{equation}
\begin{aligned}
\meanv{n_0}
&=
-\beta^{-1}\partial_{\varepsilon_0}\log\xizt,
\\
\meanv{n_0}_{N}
&=
-\beta^{-1}\partial_{\varepsilon_0}\log\znt,
\end{aligned}
\label{eq:CE-mean-n0}
\end{equation}
while the corresponding variances are
\begin{equation}
\begin{aligned}
\fluct(n_0)
&=
\beta^{-2}\partial_{\varepsilon_0}^{2}\log\xizt,
\\
\fluct_N(n_0)
&=
\beta^{-2}\partial_{\varepsilon_0}^{2}\log\znt.
\end{aligned}
\end{equation}
In the microcanonical ensemble, it is convenient to introduce the
moment-weighted thermodynamic weight
\begin{equation}
\begin{aligned}
W_N^{(m)}(E,\delta E)
&=
\sum_{\alpha:\,N_\alpha=N}
\left(n_0^{(\alpha)}\right)^m
\\
&\quad\times
\left[
\Theta(E+\delta E-E_\alpha)
-
\Theta(E-E_\alpha)
\right],
\end{aligned}
\label{eq:MCE-weighted-shell}
\end{equation}
where $n_0^{(\alpha)}$ is the ground-state occupation in the
many-body configuration $\alpha = \{n_j^{(\alpha)}\}$. For $m=0$, this reduces to the
thermodynamic weight introduced in Eq. \eqref{eq:W-Q}: $W_N^{(0)}(E,\delta E)=W_N(E,\delta E)$. Analogously to Eq.~\eqref{eq:W-Omega}, we introduce the corresponding
coarse-grained moment-weighted state density
$\Omega_N^{(m)}(E)$ through
$W_N^{(m)}(E,\delta E)\simeq\Omega_N^{(m)}(E)\delta E$.
The microcanonical moments are then
\begin{equation}
\meanv{n_0^m}_{N,E}
=
\frac{\Omega_N^{(m)}(E)}{\Omega_N(E)}.
\label{eq:MCE-moments-density}
\end{equation}
In particular, the condensate-number fluctuations are
\begin{equation}
\fluct_{N,E}(n_0)
=
\meanv{n_0^2}_{N,E}
-
\meanv{n_0}_{N,E}^{\,2}.
\label{eq:MCE-var-n0}
\end{equation}
The occupation statistics can also be related to correlation functions. %
Specifically, condensate-number fluctuations are directly related to the second-order coherence of the condensate~\cite{Wright2012}. Correlation functions also provide a natural framework for characterizing spatial coherence and density correlations~\cite{NaraschewskiGlauber1999}. For a bosonic field operator $\hat{\Psi}(\mathbf r)$, the general normally ordered $n$-th order correlation function is defined as
\begin{align}
G^{(n)}(\mathbf r_1,\ldots,&\mathbf r_n;\mathbf r'_1,\ldots,\mathbf r'_n)
=\nonumber\\
&\avggce{
\hat{\Psi}^\dagger(\mathbf r_1)\cdots \hat{\Psi}^\dagger(\mathbf r_n)
\hat{\Psi}(\mathbf r'_n)\cdots \hat{\Psi}(\mathbf r'_1)
}.
\label{eq:correlations-G}
\end{align}
The corresponding normalized correlation function is
\begin{equation}
g^{(n)}(\mathbf r_1,\ldots,\mathbf r_n;\mathbf r'_1,\ldots,\mathbf r'_n)
=
\frac{
G^{(n)}(\mathbf r_1,\ldots,\mathbf r_n;\mathbf r'_1,\ldots,\mathbf r'_n)
}{
\sqrt{
\prod_{j=1}^{n}
\rho(\mathbf r_j)\rho(\mathbf r'_j)
}
},
\label{eq:normalized-g}
\end{equation}
where
$
\rho(\mathbf r)=G^{(1)}(\mathbf r; \mathbf r)
=
\avggce{
\hat{\Psi}^\dagger(\mathbf r)\hat{\Psi}(\mathbf r)
}
$
is the local density.

The first-order correlation function measures the coherence of the matter wave, in the sense of Glauber's quantum theory of optical coherence~\cite{Glauber1963OpticalCoherence}. In the condensed phase, it is mostly determined by the mean occupation of the macroscopically populated single-particle ground state. The second-order correlation function has a different character: it is a density-density correlator and is therefore sensitive to fluctuations of the occupations, in particular to the fluctuations of the condensate population. In trapped Bose gases, such density correlations and their physical meaning were discussed in detail by Naraschewski and Glauber~\cite{NaraschewskiGlauber1999}. Spatial pair correlations in interacting one-dimensional Bose gases have also been
investigated at finite temperature ~\cite{Kheruntsyan03,Deuar09,Panfil2014}.

We would like to point out that higher order correlation functions are experimentally accessible as discussed in more detail in Sec.~\ref{Experiments}. For example, the second order correlation function was previously studied for an expanding ultracold gas, showing  bunching for a thermal gas and a constant $g^{(2)}=1$ for a BEC~\cite{Schellekens2005,Jeltes2007,Cayla2020}. Density fluctuations in an elongated Bose gas were measured indirectly by Estève~\textit{et al.}, giving indirect access to second-order density correlations through atom-number fluctuations in finite imaging cells~\cite{Esteve2006DensityFluctuations}.

\subsection{Cauchy Integral Representations for the Canonical Partition Function}

The canonical partition function can be obtained from the grand canonical one by extracting the coefficient of $z^N$ in Eq.~\eqref{eq:Xi-Zn}
through Cauchy's integral formula, as in
Refs.~\cite{Kubo1965StatMech,Ahlfors1979,HolthausKalinowski1999}.
Our construction is closest in detail to that of Holthaus and Kalinowski
\cite{HolthausKalinowski1999}
\begin{equation}
Z_N(\beta)
=
\frac{1}{2\pi i}
\oint_{\mathcal C_z}
\frac{\Xi(\beta,z)}{z^{N+1}}\,dz,
\label{eq:ZN-Cauchy}
\end{equation}
where $\mathcal C_z$ is a positively oriented contour in the complex
$z$-plane that encloses the origin while excluding all singularities
$z_j$.

We first choose a circular contour in the $z$-plane which crosses the
positive real axis at $z=z_0$, and then, via the change of variable
$w=\log z$, parametrize the contour as
\begin{equation}
w(\phi)=w_0+i\phi,
\qquad
w_0=\log z_0,
\qquad
\phi\in[-\pi,\pi). \notag
\end{equation}
In this representation Eq.~\eqref{eq:ZN-Cauchy} becomes
\begin{equation}
Z_N(\beta)
=
\frac{1}{2\pi}
\int_{-\pi}^{\pi}
e^{
\log\Xi(\beta,w_0+i\phi)
-N(w_0+i\phi)
}
\,\mathrm{d}\phi .
\label{eq:ZN-w-contour}
\end{equation}
Equivalently, we introduce the function $f(w,\beta)$, which at the saddle-point is proportional to the Helmholtz free energy:
\begin{equation}
f(w,\beta)=\log \Xi(w,\beta)-Nw,
\end{equation}
we may write
\begin{equation}
Z_N(\beta)
=
\frac{1}{2\pi}
\int_{-\pi}^{\pi}
e^{f(w_0+i\phi,\beta)}
\,\mathrm{d}\phi .
\label{eq:ZN-f}
\end{equation}

\subsection{Saddle-point approximation, its improvements and its limitations}
\label{sec:saddle-limitations}
The integral \eqref{eq:ZN-f} can be approximated using the saddle-point method, whose application to statistical partition functions dates back to Darwin and Fowler~\cite{DarwinFowler1922}. We compare the naive implementation of this technique with the more careful treatment developed by Holthaus and Kalinowski \cite{HolthausKalinowski1999}. Only a brief review of the saddle-point method is given below; for a more comprehensive treatment, see \cite{Wong1989}.

The saddle-point method is frequently used to estimate integrals of the form $\int_{C} g(w) 
e^{f(w)} \mathrm{d}w$, where the exponential factor is sharply localized near a saddle-point of $f$. The core idea is to deform the contour $C$ so that it passes through a saddle
point $w_0$, where $f'(w_0)=0$ and the real part of $f(w)$ is maximized
along the integration path. For the contour $w(\phi)=w_0+i\phi$, expansion to second order gives
\[
f(w_0+i\phi)
=
f(w_0)
-\frac{1}{2}f''(w_0)\phi^2
+\mathcal{O}(\phi^3).
\]
The stationarity condition $f'(w_0)=0$ leads to the equation
\begin{align}
\sum_{j \geq 0} \avggce{n_j} = N ,
\label{eq:conditionw0}
\end{align}
that can be solved numerically for $w_0$, related to the GCE chemical potential via $w_0 = \beta \mu$.  Since $f''(w_0)>0$, the same point is a local maximum
of $\operatorname{Re}f$ along the vertical integration contour $w=w_0+i\phi$.
The saddle-point approximation leads to
\begin{align}
Z(N,\beta)
&\simeq
\frac{1}{2\pi}
\int_{-\pi}^{\pi}
e^{f(w_0)-\frac12f''(w_0)\phi^2}
\,\mathrm{d}\phi .
\label{eq:ZNapprox}
\end{align}
In general, the Taylor series of the exponent about the saddle-point need not converge over the entire range of integration. Equation \eqref{eq:ZNapprox} remains accurate provided that the approximation fails only in the tails, where the contribution to the integral is negligible. The term $f''(w_0)\phi^2$ is of order unity when $\phi$ is of order
$[f''(w_0)]^{-1/2}$. This motivates the necessary condition for the
validity of the saddle-point approximation
\begin{equation}
\frac{f^{(n)}(w_0)}{\left[ f''(w_0) \right]^{n/2}} \ll 1, \quad n \geq 3.
\label{eq:condition1}
\end{equation}
When condition~\eqref{eq:condition1} is satisfied, the integrand in Eq.~\eqref{eq:ZN-f} is sharply peaked around $\phi=0$. Since $|\exp[f(w_0+i\phi)]|$ decreases monotonically with $|\phi|$, the regions near the boundaries of the integration interval contribute
negligibly to $Z(N,\beta)$. The integration limits can therefore be extended to infinity, yielding
\begin{align}
Z_{\rm naive}(N,\beta)
&=
\frac{e^{f(w_0)}}{2\pi}
\int_{-\infty}^{\infty}
e^{-\frac12f''(w_0)\phi^2}
\,\mathrm{d}\phi
\notag\\[4pt]
&=
\frac{e^{f(w_0)}}{\sqrt{2\pi f''(w_0)}} .
\label{eq:ZN-naive}
\end{align}
The application of formula \eqref{eq:ZN-naive}, as presented for instance in \cite{HolthausKalinowski1999}, leads to nonphysical condensate fluctuations with a leading $N^2$ scaling for $T\to 0$. In the case of a three-dimensional harmonic trap, such a naive approximation would give $\fluct{ n_0} \simeq 0.02438\,N^2 + 0.01304\,N$. The reason behind it is that the condition \eqref{eq:condition1} cannot be satisfied, because $f^{(n)}(w_0)/\left[ f''(w_0) \right]^{n/2}$ is of order unity. 

The saddle-point approximation can, however, be significantly improved.
As discussed in Refs.~\cite{Dingle1973,HolthausKalinowski1999},
its failure in the condensed regime originates from the ground-state
contribution, encoded in the factor $1/(1-z)=1/(1-e^w)$ of the
grand canonical partition function $\Xi$. Following Dingle~\cite{Dingle1973}, we exclude the ground-state contribution from the part of the exponent treated within the saddle-point approximation, retaining it exactly, and expand the remaining excited-state contribution about the same point $w_0$. To this end,
we introduce $\Xi_{\rm ex}$, the contribution of the excited states to
the grand canonical partition function, and write
\begin{equation}
\Xi(w,\beta)
=
\frac{\Xi_{\rm ex}(w,\beta)}{1-e^w}
=
\Xi_{\rm ex}(w,\beta)
\sum_{n=0}^{\infty}e^{nw}. \notag
\end{equation}
Although the geometric series extends to infinity, extraction of the coefficient of $z^N$ through Cauchy's integral formula reduces the contributing terms to $0\leq n\leq N$. We denote the resulting approximation by $Z_{\rm better}$. Repeating the
steps used to obtain $Z_{\rm naive}$ and expanding the excited-state
exponent to second order around the common point $w_0$, we obtain
\begin{equation}
\begin{split}
Z_{\rm better}(N,\beta)
&=
\frac{1}{2\pi}
\sum_{n=0}^{N}
\int_{-\infty}^{\infty}
\exp\!\Bigl[
f_{\rm ex}(w_0,n)
\\
&\quad
+i f_{\rm ex}'(w_0,n)\phi
-\frac12 f_{\rm ex}''(w_0)\phi^2
\Bigr]
\,\mathrm{d}\phi .
\end{split}
\end{equation}
where
\begin{equation}
f_{\rm ex}(w,n)
=
\log \Xi_{\rm ex}(w,\beta)-(N-n)w.
\label{eq:fex}
\end{equation}
The expansion point $w_0$ is determined by
Eq.~\eqref{eq:conditionw0}. Since
\begin{equation}
\left.
\partial_w\log\Xi_{\rm ex}(w,\beta)
\right|_{w=w_0}
=
N-\avggce{n_0}, \notag
\end{equation}
the linear coefficient is
\begin{equation}
f_{\rm ex}'(w_0,n)
=
n-\avggce{n_0}. \notag
\end{equation}
Thus, only the term with $n=\avggce{n_0}$ is stationary at $w_0$; for the remaining terms, the linear phase must be retained. The Gaussian integral can then be evaluated analytically, yielding
\begin{equation}
Z_{\rm better}(N,\beta)
\simeq
\sum_{n=0}^{N}
\frac{
\exp\!\left[
f_{\rm ex}(w_0,n)
-
\frac{
\left(n-\avggce{n_0}\right)^2
}{
2f_{\rm ex}''(w_0)
}
\right]
}{
\sqrt{2\pi f_{\rm ex}''(w_0)}
}.
\label{eq:ZN-pole-separated}
\end{equation}

\begin{figure}
    \centering
        \includegraphics[width=0.95\linewidth]{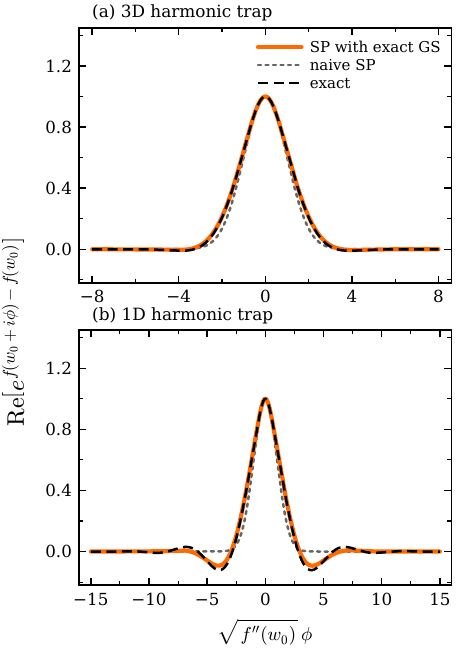}
\caption{
Comparison of the real part of the integrand in Eq.~\eqref{eq:ZN-f} with two saddle-point approximations. The black dashed curve denotes the exact result, the orange solid curve
the saddle-point approximation with the ground-state contribution retained exactly
(\emph{SP with exact GS}), and the grey dotted curve the naive
saddle-point approximation.
(a) Three-dimensional harmonic trap with $N=500$ and
$T/T_{\mathrm c}=0.9$, where
$ T_{\rm c}=(\hbar\omega/k_{\rm B})
[N/\zeta(3)]^{1/3}$.
(b) One-dimensional harmonic trap with $N=1000$ and
$T/T_{\mathrm c}=1.2$, where $T_{\mathrm c}$ is defined by
$\langle n_0\rangle(T_{\mathrm c})=N/2$.
}
\label{fig:saddle-comparison}
\end{figure}

To illustrate the improvement over the naive saddle-point approximation,
we compare the real part of the ratio of the exact integrand in
Eq.~\eqref{eq:ZN-f} to its value at the saddle point,
\begin{equation}
\operatorname{Re}\!\left[
\exp\!\left(f(w_0+i\phi)-f(w_0)\right)
\right], \notag
\end{equation}
with the naive Gaussian approximation obtained by expanding the exponent up to second order, and with the corresponding approximation in which
the ground-state contribution is retained exactly and only the excited-state
part is expanded to second order. These quantities are shown
as functions of $\sqrt{f''(w_0)}\,\phi$ in Fig.~\ref{fig:saddle-comparison}.

Here we argue that such an approach, although significantly better than the naive approach, would still lead in some cases to systematic deviations, even in the thermodynamic limit. The reason is that the radius of convergence of the Taylor series expansion $f_{\rm ex}(w)$ can often be much smaller than the range of integration. In some cases, as $\phi$ approaches the radius of convergence, the integrand becomes negligible. This happens, for instance, for a Bose gas trapped in a three-dimensional harmonic trap approaching the thermodynamic limit. Otherwise, the Taylor expansion is no longer valid on a substantial part of the integration range, and saddle-point approximations are no longer accurate. We refer to systems exhibiting this behaviour as non-Gaussian. The categorization into Gaussian and non-Gaussian systems can be conveniently described using cumulants. We introduce them briefly below, as they will play an important role in our method.

\subsection{Gaussian systems and cumulants}
\label{sec:gaussianity}
Cumulants provide a convenient measure of non-Gaussianity of a probability distribution \cite{Kocharovsky2000,KocharovskyReview2006}. We introduce the mixed grand canonical cumulants of $N_{\rm ex}$ and the dimensionless energy $U$, defined at the real saddle point by
\begin{equation}
\kappa(\nu_1,\nu_2)
=
(-1)^{\nu_2} \,
\frac{\partial^{\nu_1+\nu_2}\ln \Xi_{\rm ex}(w_0,b_0)}
{\partial w^{\nu_1}\partial b^{\nu_2}},
\label{eq:cumulants_definition}
\end{equation}
where we have introduced the dimensionless variables
\begin{equation}
w_0=\beta\mu,\qquad
b_0=\frac{\delta\varepsilon}{k_{\rm B}T},
\qquad
U=\frac{E}{\delta\varepsilon},
\label{eq:dimensinless-vars}
\end{equation}
and let $\delta\varepsilon$ be the lowest single-particle excitation energy,
\begin{equation}
\delta\varepsilon=\varepsilon_1-\varepsilon_0.
\end{equation}
Here $(w_0,b_0)$ denote the real saddle-point values corresponding to $(N,U)$.
We use the same notation for functions expressed in these
dimensionless variables; in particular, $Z_N(b)$ and $\Omega_N(U)$.

For brevity, we also introduce the following notation for cumulants:
\begin{equation}
\label{eq:cumulant-definitions}
\begin{aligned}
\kappa_{ \nu_1 \nu_2}&= \kappa(\nu_1,\nu_2), &
\kappa_{\nu_1} &= \kappa(\nu_1,0).
\end{aligned}
\end{equation}
The cumulants can be easily expressed in terms of the average grand canonical occupations of the $i$-th excited state, $\avggce{ n_i } $. For instance:
\begin{align}
\kappa_{1}
&=
\sum_{i\geq 1}
\avggce{n_i} 
=
\avggce{N_{\mathrm{ex}}}, \notag
\\
\kappa_{2}
&=
\sum_{i\geq 1}
\left(
\avggce{n_i^2}
-
\avggce{n_i}^{\,2}
\right)
=
\fluct{N_{\mathrm{ex}}}, \notag
\\
\kappa_{01}
&=
\sum_{i\geq 1}
\frac{\varepsilon_i}{\delta\varepsilon}
\avggce{n_i}
=U. \notag
\end{align}

More convenient measures of statistical properties of a distribution approaching the thermodynamic limit are standardized cumulants, $ \gamma_{\nu_1 \nu_2}$ as they remain finite with the growing size of a system \cite{kubo1962, KenneyKeeping1962}. We define 
\begin{align}
    \gamma_{\nu_1 \nu_2} = \frac{\kappa_{\nu_1 \nu_2}}
    {\kappa_{2}^{\nu_1/2} \kappa_{02}^{\nu_2/2}},
\label{eq:standardized-cumulants}
\end{align}
where, for brevity, we use the notation $\gamma_m$ for $\gamma_{m0}$. We call a system Gaussian if
\begin{equation}
\gamma_m \xrightarrow[\scriptscriptstyle\mathrm{TL}]{}0,
\qquad
\gamma_{0m} \xrightarrow[\scriptscriptstyle\mathrm{TL}]{}0,
\qquad m\geq3 .
\label{eq:condition_gaussianity}
\end{equation}
For the ideal Bose gas, H\"older's inequality gives, with
$m=\nu_1+\nu_2$,
\begin{equation}
\gamma_{\nu_1\nu_2}
\leq
\gamma_m^{\nu_1/m}
\gamma_{0m}^{\nu_2/m}.
\label{eq:mixed-cumulant-holder}
\end{equation}
Hence Eq.~\eqref{eq:condition_gaussianity} implies
$\gamma_{\nu_1\nu_2}\to0$ for all $\nu_1+\nu_2\geq3$.
This condition is satisfied, in particular, for a three-dimensional
harmonic trap \cite{Kocharovsky2000}.

\section{Main Results}
\subsection{Segmented Contour Method: Statistics in Canonical Ensemble
\label{subsec:segmented-ZN}}
In this subsection, we introduce the segmented-contour method, which extends the approaches of Refs.~\cite{Dingle1973,HolthausKalinowski1999} described above to properly account for non-Gaussianity and allow the statistics to be computed accurately and efficiently for large systems. We introduce the rescaled contour coordinate
\begin{equation}
t=\sqrt{\kappa_2}\,\phi,
\qquad
w(t)=w_0+i\frac{t}{\sqrt{\kappa_2}},
\end{equation}
with
\begin{equation}
\mathcal I_t
=
\left[-\pi\sqrt{\kappa_2},\,\pi\sqrt{\kappa_2}\right].
\label{eq:canonical-contour-domain}
\end{equation}
In terms of the new integration variable, Eq.~\eqref{eq:ZN-f} becomes
\begin{equation}
Z_N(\beta)
=
\frac{1}{2\pi\sqrt{\kappa_2}}
\int_{\mathcal I_t}
e^{f(w(t),\beta)}
\,\mathrm{d}t
\label{eq:ZN-rescaled}
\end{equation}
can be approximated by dividing the contour integral into a finite number of  intervals and applying Gaussian approximation on each segment.
In this way, it accounts for non-Gaussian corrections, while preserving the low numerical cost of saddle-point methods. 
For Gaussian systems, a single segment often suffices, and the method reduces to the standard saddle-point approximation.

We choose $\Delta_t$ such that
\begin{equation}
K_t = \frac{\sqrt{\kappa_2}\pi}{\Delta_t}
\end{equation}
is a positive integer, and divide the positive half of the integration
interval, denoted by $\mathcal I_t^{+}$, into $K_t$ sufficiently short
subintervals of length $\Delta_t$,
\begin{equation}
\mathcal I_t^+
=
\bigcup_{j=0}^{K_t-1}\mathcal T_j,
\qquad
\mathcal T_j=[t_j,t_{j+1}),
\qquad
t_j=j\Delta_t,
\label{eq:contour-partition}
\end{equation}
on each of which the exponent can be approximated up to second order. Since the contribution from $t<0$ is the complex conjugate of the
contribution from $t>0$, the integral in Eq.~\eqref{eq:ZN-f} can be
written as twice the real part of the positive-$t$ contribution
\begin{equation}
\begin{aligned}
Z_N &= 2\,\operatorname{Re} I_+,
\\
I_+
&=
\frac{1}{2\pi\sqrt{\kappa_2}}
\sum_{n=0}^{N}
\sum_{j=0}^{K_t-1}
\int_{\mathcal T_j}
e^{f_{\rm ex}(w(t),n)}
\,\mathrm{d}t .
\end{aligned}
\label{eq:Iplus-full}
\end{equation}
The segment width $\Delta_t$ is chosen sufficiently small that
$f_{\rm ex}$ has a convergent local Taylor expansion on each
$\mathcal T_j$. From the total number of segments, we retain
$M_t<K_t$ segments whose contribution to the integral is non-negligible.
The choice of the parameters $M_t$ and $\Delta_t$ controls both the
accuracy and the computational cost of the calculation. The expression
above then becomes
\begin{align}
I_+ \simeq
\frac{1}{2\pi\sqrt{\kappa_2}}
\sum_{n=0}^{N}
\sum_{j=0}^{M_t-1}
\int_{\mathcal T_j}
e^{f_{\rm ex}(w(t),n)}
\,\mathrm{d}t .
\notag
\end{align}

\begin{figure}
    \centering
    \includegraphics[
        width=0.95\linewidth
    ]{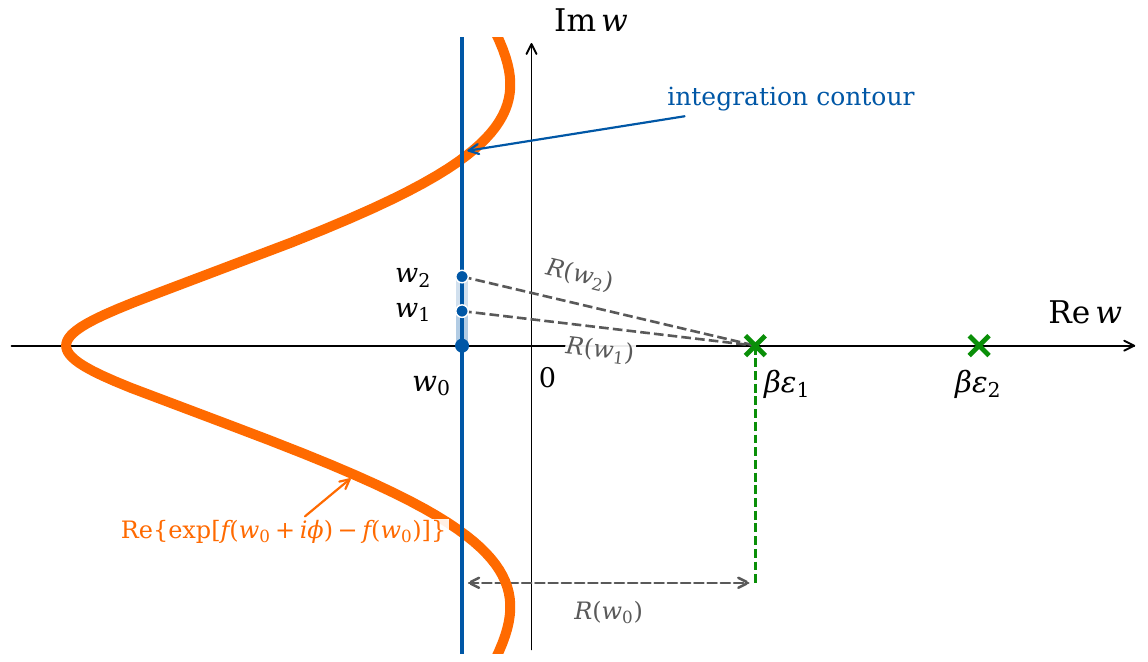}
    \caption{Schematic representation of contour integration of function $\mathrm{Re} \lbrace \exp\left[ f(w) - f(w_0)\right] \rbrace$ (orange line) from Fig. \ref{fig:saddle-comparison}. The contour (blue line) is divided into segments of length smaller than $R(w_0)$, which ensures the convergence of Taylor expansion at each $w_j$ with $j \geq 0 $.
    }
    \label{fig:complex-w-contour}
\end{figure}

For the excited-state grand canonical partition function,
$\Xi_{\mathrm{ex}}$, the singularities are located at
$w=\beta\varepsilon_j+2\pi i k$, with
$j\geq1$ and $k\in\mathbb Z$.
The Taylor series of $\log\Xi_{\mathrm{ex}}(w)$ about $w_0$
therefore converges in the largest disk centred at $w_0$
that contains no such singularities. At the partition points introduced above, we write
$w_j\equiv w(t_j)$. The corresponding radii of convergence
are denoted by $R(w_j)$. In particular, $R(w_0)$ is smaller
than the radii $R(w_1), \, R(w_2), \, \ldots$ associated with the
subsequent partition points, as illustrated in
Fig.~\ref{fig:complex-w-contour}.
For the rescaled contour coordinate $t$, the radius of convergence
about any partition point $w_j=w(t_j)$ is bounded from below by
\begin{equation}
\Delta_R
=
\sqrt{\kappa_2}\,R(w_0)
=
\sqrt{\kappa_2}(b_0-w_0),
\label{eq:Radiusconvergence}
\end{equation}
where $b_0$ is defined in Eq.~\eqref{eq:dimensinless-vars} and
$\varepsilon_0=0$. Hence, choosing $\Delta_t<\Delta_R$ guarantees
convergence of the local Taylor expansion on every segment
$\mathcal T_j$. On each of the $M_t$ retained segments, we expand
$f_{\rm ex}(w(t),n)$ about $w_j=w(t_j)$ in the local variable
$t-t_j$ and truncate the expansion at second order. The corresponding
piecewise-quadratic approximation to the integrand is denoted by
$e^{\tilde f_{\rm ex}(t,n)}$ and is set to zero outside the retained
segments. Its explicit construction is given in
Appendix~\ref{app:segmented-method}. The canonical partition function
is then approximated as
\begin{equation}
\begin{aligned}
Z_N &= 2\,\operatorname{Re} I_+, \\
I_+ &\simeq
\frac{1}{2\pi\sqrt{\kappa_2}}
\sum_{n=0}^{N}
\int_{\mathcal I_t^+}
e^{\tilde f_{\rm ex}(t,n)}
\,\mathrm dt .
\end{aligned}
\label{eq:ZNfinal}
\end{equation}

The remaining difficulty is to choose $\Delta_t$ and $M_t$ such that the
local quadratic approximation and the truncation of the contour are
both accurate. The segment width $\Delta_t$ must remain within the convergence domain
of the local Taylor expansion. In addition, the leading correction to
the quadratic exponent must be negligible. Since the cubic term is
bounded by $\gamma_3\Delta_t^3/6$, we require
\begin{equation}
\Delta_t<\Delta_R,
\qquad
\frac{\gamma_3}{6}\Delta_t^3\ll1.
\label{eq:conditionsDelta}
\end{equation}
The standardized cumulants evaluated at fixed $t$ on the line $w(t)$ approach finite limiting functions as the system size increases. This indicates that the  integrand in the variable $t$ remains on an $\mathcal O(1)$ scale, while the full integration interval grows as $\pi\sqrt{\kappa_2}$. Hence, at fixed numerical accuracy, the retained range can be chosen such that
\begin{equation}
1 \ll M_t \, \Delta_t\ll \pi\sqrt{\kappa_2}.
\label{eq:Mtildeconstraint}
\end{equation}
Thus, provided an $\mathcal O(1)$ segment width satisfying condition~\eqref{eq:Mtildeconstraint} can be chosen, an $\mathcal O(1)$ number of segments is sufficient and $M_t$ need not grow with the system size.

To calculate the moments of $n_0$, we notice that each term in the sum over $n$ corresponds to a fixed condensate occupation $n_0=n$, hence, the moments of $n_0$ are obtained by weighting the corresponding terms in Eq.~\eqref{eq:ZNfinal}
by $n^m$:
\begin{equation}
\begin{aligned}
\langle n_0^m\rangle_N
&=
\frac{\operatorname{Re} I_+^{(m)}}
{\operatorname{Re} I_+^{(0)}},
\\
I_+^{(m)}
&\simeq
\frac{1}{2\pi\sqrt{\kappa_2}}
\sum_{n=0}^{N}
\int_{\mathcal I_t^+}
n^m\,
e^{\tilde f_{\rm ex}(t,n)}
\,\mathrm{d}t .
\end{aligned}
\label{eq:canonical-moments}
\end{equation}
\begin{figure}
\includegraphics[width=0.95\linewidth]{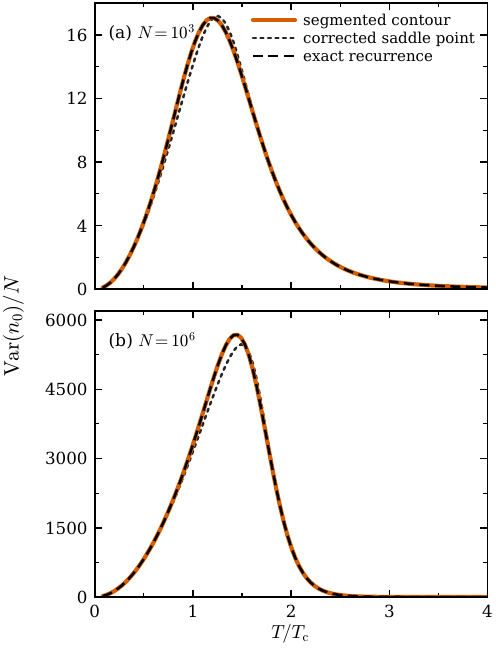}
\caption{
Canonical condensate-number fluctuations of a one-dimensional harmonically trapped ideal Bose gas.
We plot $\fluct_N(n_0)/N$ as a function of $T/T_{\mathrm c}$, with $\omega=D/N$ and $D=100$.
The segmented-contour result (thick orange solid line) is compared with the corrected saddle-point approximation of Refs.~\cite{Dingle1973,HolthausKalinowski1999} (black dotted line) and with the exact recurrence result (black dashed line).
Panels (a) and (b) correspond to $N=10^3$ and $N=10^6$, respectively.
For the temperature range shown, the retained numbers of contour segments and their widths were
$M_t=52$--$82$, $\Delta_t=0.314$--$0.393$ for $N=10^3$, and
$M_t=66$--$82$, $\Delta_t=0.314$--$0.393$ for $N=10^6$.
}
\label{fig:fluctuations_compare}
\end{figure}
Figure~\ref{fig:fluctuations_compare} illustrates the performance of the
segmented-contour method for a one-dimensional harmonically trapped ideal
Bose gas. For $N=10^3$, we compare the segmented-contour result with the
exact canonical recurrence of Eq.~\eqref{eq:canonical-recurrence} and with
the corrected saddle-point approximation of
Refs.~\cite{Dingle1973,HolthausKalinowski1999}. For $N=10^6$, where the
exact recurrence calculation becomes computationally demanding, we compare
the segmented-contour result with the corrected saddle-point approximation. By the criterion in Eq.~\eqref{eq:condition_gaussianity}, the one-dimensional harmonic trap is non-Gaussian \cite{Kocharovsky2000}. Consequently, the single-saddle approximation
retains a systematic error even at large system size, as seen in
Fig.~\ref{fig:fluctuations_compare}.

\subsection{Statistics in the Microcanonical Ensemble \label{subsec:microocanonical}}

We now extend our approach to the microcanonical
ensemble, where both $N$ and $U$ are fixed.
Since the canonical
partition function is its Laplace transform, the energy constraint can
be imposed by the inverse Laplace transform \cite{Kubo1965StatMech},
\begin{equation}
\Omega_N(U)
=
\frac{1}{2\pi i}
\int_{b_0-i\infty}^{b_0+i\infty}
Z_N(b)\,e^{bU}\,\mathrm db.
\label{eq:Omega-inverse-Laplace}
\end{equation}
We do not require the single-particle spectrum to be commensurate, as defined in Sec.~\ref{sec:System}. To this end, we introduce the auxiliary function
\begin{equation}
s_{\rm ex}(w,b,n)
=
\log \Xi_{\rm ex}(w,b)
-(N-n)w+b \,U .
\label{eq:sex}
\end{equation}

We parametrize the particle-number and energy contours as
\begin{equation}
w(t)=w_0+i\frac{t}{\sqrt{\kappa_2}},
\qquad
b(s)=b_0+i\frac{s}{\sqrt{\kappa_{02}}}.
\label{eq:MCE-contours}
\end{equation}
In contrast to the canonical case, the integration domain associated with the inverse-Laplace energy variable is unbounded. Hence, we define
\begin{equation}
\mathcal I_s=\mathbb R,
\qquad
\mathcal D=\mathcal I_t^+\times\mathbb R .
\label{eq}
\end{equation}
Using the symmetry under $(t,s)\mapsto(-t,-s)$, the microcanonical
state density can then be written as
\begin{align}
\Omega_N(U)
&=
\frac{2}{(2\pi)^2\sqrt{\kappa_2\kappa_{02}}}
\operatorname{Re}
\sum_{n=0}^{N}
\int_{\mathcal D}
e^{s_{\rm ex}(w(t),b(s),n)}
\,\mathrm dt\,\mathrm ds .
\label{eq:Omegasex}
\end{align}
For a given number of atoms $N$ and dimensionless energy $U = E/\delta\varepsilon$, the saddle-point $(w_0, b_0)$ is determined by the conditions
\begin{equation}
\begin{aligned}
\partial_w s_{\rm ex}(w_0,b_0,\langle n_0\rangle)
&=
\kappa_1+\langle n_0\rangle-N=0,
\\
\partial_b s_{\rm ex}(w_0,b_0,\langle n_0\rangle)
&=
-\kappa_{01}+U=0.
\end{aligned}
\label{eq:saddleMCE}
\end{equation}

Analogously to the canonical case of
Sec.~\ref{subsec:segmented-ZN}, the particle-number contour is
partitioned into the intervals $\mathcal T_j$ defined in
Eq.~\eqref{eq:contour-partition}, while the energy contour is divided
into intervals of width $\Delta_s$,
\begin{equation}
\mathcal S_k=[s_k,s_{k+1}),
\qquad
s_k=k\Delta_s,
\qquad
k\in\mathbb Z .
\label{eq:energy-contour-partition}
\end{equation}
The segment widths $\Delta_t$ and $\Delta_s$ control the accuracy of
the local quadratic approximation. A sufficient condition for the
local Taylor expansions to remain within their domains of analyticity is
\begin{equation}
\frac{\Delta_s}{\sqrt{\kappa_{02}}}<b_0,
\qquad
\frac{\Delta_t}{\sqrt{\kappa_2}}
+
\frac{\Delta_s}{\sqrt{\kappa_{02}}}
<
b_0-w_0.
\label{eq:MCE-analyticity-condition}
\end{equation}

For the ideal Bose gas, the cumulants evaluated along the vertical
contours are bounded in absolute value by their values at the saddle point, as shown in Appendix~\ref{app:segmented-method}. Combining this
with Eq.~\eqref{eq:mixed-cumulant-holder} for $m=3$, the complete
third-order contribution in the local rescaled coordinates
$\phi=t-t_j$ and $\theta=s-s_k$ obeys
\begin{equation}
|R_3(\phi,\theta)|
\leq
\frac{1}{6}
\left(
\gamma_3^{1/3}|\phi|
+
\gamma_{03}^{1/3}|\theta|
\right)^3.
\label{eq:MCE-third-order-bound}
\end{equation}
A sufficient condition for the leading  truncation error to be negligible on every retained subdomain is therefore
\begin{equation}
\frac{1}{6}
\left(
\gamma_3^{1/3}\Delta_t
+
\gamma_{03}^{1/3}\Delta_s
\right)^3
\ll 1.
\label{eq:MCE-cubic-condition}
\end{equation}
The particle-number contour is treated as in the canonical construction, with
\begin{equation}
1\ll M_t\Delta_t\ll\pi\sqrt{\kappa_2}.
\label{eq:MCE-number-tail-condition}
\end{equation}
For the unbounded energy contour, we retain
\begin{equation}
k=-M_s,\ldots,M_s-1,
\qquad
S_s=M_s\Delta_s,
\label{eq:MCE-energy-range}
\end{equation}
where the retained range is chosen to contain all appreciable contributions. 

\begin{figure}
\centering
    \includegraphics[width=0.95\linewidth]{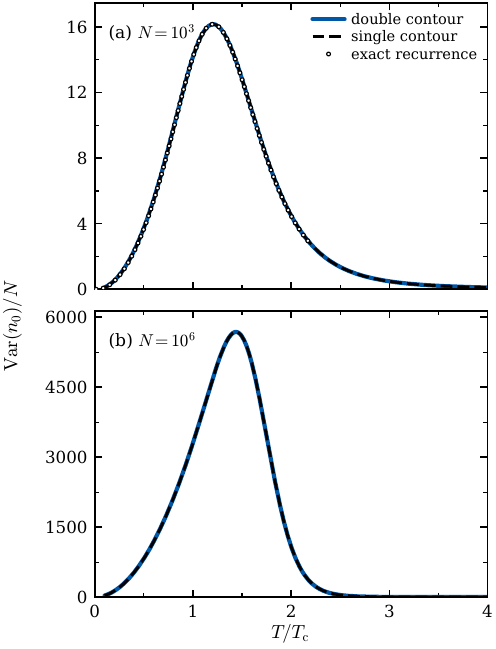}
    \caption{
Microcanonical condensate-number fluctuations of a one-dimensional
harmonically trapped ideal Bose gas.
We plot $\fluct_{N,E}(n_0)/N$ as a function of $T/T_{\mathrm c}$,
with $\omega=D/N$ and $D=100$.
The solid blue curve shows the full segmented double-contour result,
and the black dashed curve the single-contour approximation.
In panel (a), open circles show the exact microcanonical recurrence
benchmark obtained from Eq.~\eqref{eq:microcanonical-recurrence}.
(a) $N=10^3$; (b) $N=10^6$.
}
\label{fig:comparison}
\end{figure}

On each retained subdomain
$\mathcal T_j\times\mathcal S_k$, we expand
$s_{\rm ex}(w(t),b(s),n)$ about the corresponding partition point
$(w(t_j),b(s_k))$ and truncate the bivariate Taylor expansion at
second order. The resulting piecewise-quadratic approximation is denoted
by $e^{\tilde s_{\rm ex}(t,s,n)}$ and is set to zero outside the retained
subdomains. Its explicit construction is given in
Appendix~\ref{app:segmented-method}. The microcanonical state density
and the moments of $n_0$ can then be written as
\begin{equation}
\begin{aligned}
\Omega_N(U)
&=
2\,\operatorname{Re}I_+^{(0)},
\qquad
\langle n_0^m\rangle_{N,E}
=
\frac{\operatorname{Re}I_+^{(m)}}
{\operatorname{Re}I_+^{(0)}},
\\
I_+^{(m)}
&\simeq
\frac{1}{(2\pi)^2\sqrt{\kappa_2\kappa_{02}}}
\sum_{n=0}^{N}
\int_{\mathcal D}
n^m e^{\tilde s_{\rm ex}(t,s,n)}
\,\mathrm dt\,\mathrm ds .
\end{aligned}
\label{eq:OmegaNfinal}
\end{equation}
Here $m\geq0$, with $m=0$ corresponding to the microcanonical
state density $\Omega_N$. For comparison, we also consider a reduced form of
Eq.~\eqref{eq:OmegaNfinal}; the corresponding results are shown in
Fig.~\ref{fig:comparison}. In many cases, $M_s$ can be set to unity, so that the integration over $s$ in Eq.~\eqref{eq:OmegaNfinal} reduces to a Gaussian integral that can be evaluated analytically. The remaining particle-number contour may still require segmentation. This approach applies for a broad class of sufficiently large systems, including ideal Bose gases with power-law spectra, as discussed in Appendix~\ref{app:segmented-method}.

\subsection{Gaussian limit}
\label{subsec:gaussian-limit}
We now consider Gaussian systems, defined by the condition \eqref{eq:condition_gaussianity}. 
For instance, for an ideal Bose gas with a power-law single-particle spectrum $\varepsilon_{\mathbf q}\propto|\mathbf q|^\sigma$ in a trap of dimension $d$, this condition is satisfied for $d \geq 2\sigma$, whereas systems with $d<2\sigma$ remain non-Gaussian \cite{KocharovskyReview2006}.  For a sufficiently large Gaussian system, the appreciable contribution
to the contour integrals in
Eqs.~\eqref{eq:canonical-moments} and \eqref{eq:OmegaNfinal}
is confined to an $\mathcal O(1)$ neighborhood of the real saddle point. This region can be covered by a single central segment in the canonical case and by the two subdomains adjacent to $s=0$ in the microcanonical case, covering the entire domain. We thus set $M_t=1$ in the CE and $M_t=M_s=1$ in the MCE, and extend $\Delta_t$ and $\Delta_s$ to infinity. The aim of this section is to derive simple analytical formulas that are exact for Gaussian systems in the thermodynamic limit and remain good approximations for finite system sizes.

\begin{figure}[t]
    \centering

    \begin{tikzpicture}
        \node[inner sep=0pt] (img) {
            \includegraphics[width=\columnwidth]
            {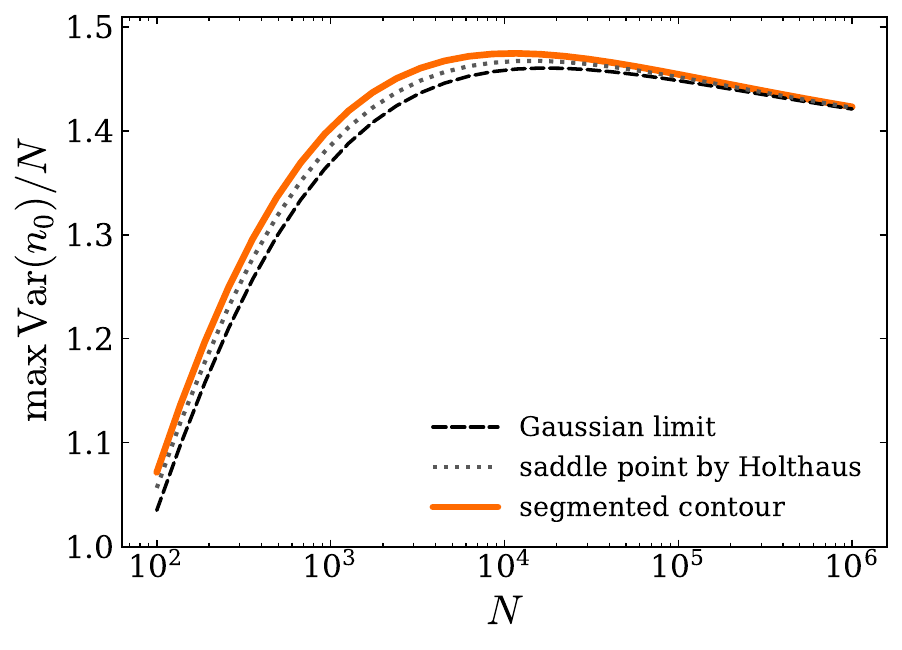}
        };
        \node[
            anchor=north east,
            xshift=-8mm,
            yshift=-4mm,
            font=\bfseries
        ] at (img.north east) {(a)};
    \end{tikzpicture}

    \par\vspace{0.3em}

    \begin{tikzpicture}
        \node[inner sep=0pt] (img) {
            \includegraphics[width=\columnwidth]
            {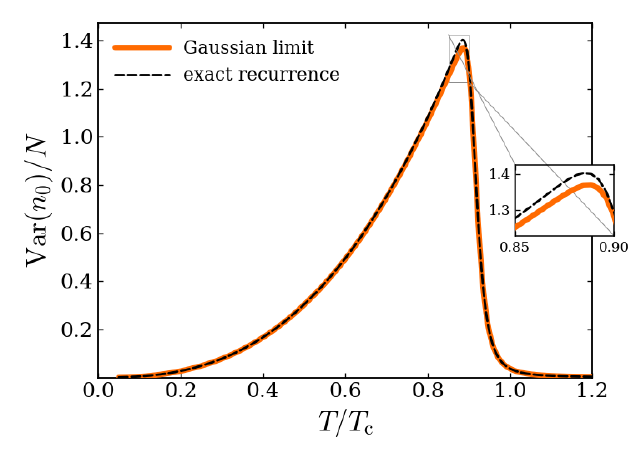}
        };
        \node[
            anchor=north east,
            xshift=-8mm,
            yshift=-4mm,
            font=\bfseries
        ] at (img.north east) {(b)};
    \end{tikzpicture}

   \caption{ %
Gaussian-limit benchmark for a three-dimensional harmonic trap.
(a) Maximum of $(\fluct n_0)_N/N$ as a function of particle number.
The segmented-contour result is shown by the orange solid curve, the
Gaussian-limit approximation by the black dashed curve, and the saddle-point approximation of Ref.~\cite{HolthausKalinowski1999} by the grey dotted curve.
(b) Condensate-occupation fluctuations for $N=1000$.
The orange solid curve shows the Gaussian-limit result, while the black
dashed curve denotes the exact recurrence benchmark.
The inset highlights the difference near the maximum.
}
    \label{fig:gaussian-limit-benchmark}
\end{figure}

Performing the integration in Eq.~\eqref{eq:canonical-moments} then yields
\begin{align}
\langle n_0^m\rangle_N
&\simeq
\frac{
\displaystyle
\sum_{n=0}^{N}
n^m
\exp\!\left[
n w_0
-\frac{(n-\avggce{n_0})^2}{2\kappa_2}
\right]
}{
\displaystyle
\sum_{n=0}^{N}
\exp\!\left[
n w_0
-\frac{(n-\avggce{n_0})^2}{2\kappa_2}
\right]
}.
\label{eq:gaussian-moments-CE}
\end{align}
Since the moments of a discrete probability distribution satisfy
$\langle n_0^m\rangle_N=\sum_{n=0}^{N}n^m P_N(n_0=n)$,
comparison with Eq.~\eqref{eq:gaussian-moments-CE} gives
\begin{equation}
\begin{aligned}
P_N(n_0=n)
&\propto
\exp\!\left[
n w_0
-
\frac{(n-\avggce{n_0})^2}{2\kappa_2}
\right],
\\
&\qquad n=0,\ldots,N.
\end{aligned}
\label{eq:Pn0_CE_gaussian}
\end{equation}
where $\avggce{n_0}$ is the grand canonical mean condensate
occupation defined in Eq.~\eqref{eq:CE-mean-n0} evaluated at $w_0$. For the microcanonical ensemble, performing the Gaussian integrations
in Eq.~\eqref{eq:OmegaNfinal} analogously gives
\begin{equation}
\begin{aligned}
P_{N,E}(n_0=n)
&\propto
\exp\!\left[
n w_0
-
\frac{(n-\avggce{n_0})^2}
{2\left(\kappa_2-\kappa_{11}^2/\kappa_{02}\right)}
\right],
\\
&\qquad n=0,\ldots,N.
\end{aligned}
\label{eq:Pn0_MCE_gaussian}
\end{equation}
In Eqs.~\eqref{eq:Pn0_CE_gaussian} and
\eqref{eq:Pn0_MCE_gaussian}, the proportionality constants are fixed
by normalization, $\sum_{n=0}^{N}P_X(n_0=n)=1$, and the
distributions vanish outside $0\leq n\leq N$.
Writing $X=N$ or $X=(N,E)$, we can evaluate all the moments
\begin{equation}
\langle n_0^m\rangle_X
=
\sum_{n=0}^{N}n^m P_X(n_0=n).
\label{eq:gaussian_moments_CE_MCE}
\end{equation}
The numerical benchmark is shown in Fig.~\ref{fig:gaussian-limit-benchmark}.

This result is consistent with the finite negative-binomial approximation of Ref.~\cite{KocharovskyReview2006} for the distribution of the
non-condensed occupation $N_{\rm ex}=N-n_0$. In the thermodynamic limit, this distribution approaches a Gaussian with mean
$H$ and variance $H(1+\eta)$. Since, in the condensate regime, $w_0\to0$ in the thermodynamic limit, the corresponding cumulants satisfy
\begin{equation}
H=\kappa_1,
\qquad
H(1+\eta)=\kappa_2 . \notag
\end{equation}
Using $n_0=N-N_{\rm ex}$ then reproduces
Eq.~\eqref{eq:Pn0_CE_gaussian}. The higher-cumulant corrections retained
by the finite negative-binomial approximation lie beyond the Gaussian approximation considered here.

\subsection{The ratio of maxima of fluctuations in CE and MCE}
\begin{figure}
\centering
\includegraphics[width=0.95\linewidth]{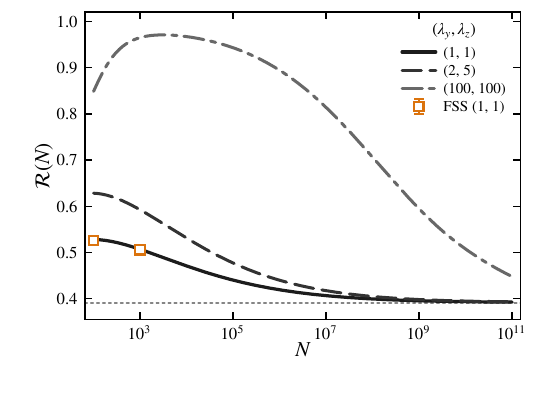}

\caption{
Ratio $\mathcal{R}(N)$ as a function of the particle number $N$ for three-dimensional anisotropic harmonic traps.
The grey dotted line denotes the thermodynamic-limit prediction
$\mathcal{R}_{\infty}=1-3\zeta(3)^2/[4\zeta(2)\zeta(4)]\simeq 0.3913$~\cite{Idziaszek2005}.
The solid, dashed, and dash-dotted curves correspond to trap anisotropies
$(\lambda_y,\lambda_z)=(1,1)$, $(2,5)$, and $(100,100)$, respectively.
Orange open squares show the revised Fock-state-sampling (FSS) results for the isotropic trap, obtained using an updated implementation of the method of Ref.~\cite{Kruk23}.
The two FSS estimates are $\mathcal{R}(100)=0.5258\pm0.0038$ and $\mathcal{R}(1000)=0.5061\pm0.0025$, with error bars indicating the reported numerical uncertainties.
Both estimates agree with the segmented-contour predictions within the reported uncertainties.
The finite-size corrections depend strongly on the trap anisotropy, whereas all curves approach the same asymptotic value for increasing $N$.
}
\label{fig:R_vs_N_aspect_ratios}
\end{figure}

We now analyze the ratio of the maxima of the condensate-number fluctuations in the microcanonical and canonical ensembles, 
\begin{equation}
\mathcal{R}(N)
=
\frac{
\displaystyle \max_E \operatorname{Var}_{N,E}(n_0)
}{
\displaystyle \max_T \operatorname{Var}_{N}(n_0)
}.
\label{eq:R}
\end{equation}
We focus on the experimentally motivated system of bosonic atoms confined in a
three-dimensional harmonic trap,
\begin{equation}
V(\mathbf{r}) =
\frac{1}{2}m\left(
\omega_x^2 x^2+\omega_y^2 y^2+\omega_z^2 z^2
\right).
\end{equation}
Different trap aspect ratios are characterized by the dimensionless parameters
\begin{equation}
\lambda_y=\frac{\omega_y}{\omega_x},
\qquad
\lambda_z=\frac{\omega_z}{\omega_x}.
\end{equation}
These parameters determine the geometry of the confinement: $\lambda_y=\lambda_z=1$ corresponds to an isotropic harmonic trap, whereas $\lambda_y,\lambda_z\gg1$ correspond to a cigar-shaped trap elongated along the $x$ direction, and $\lambda_y,\lambda_z\ll1$ to a pancake-shaped trap tightly confined along $x$. Such anisotropic configurations are particularly relevant for studying the crossover towards lower-dimensional regimes. Figure~\ref{fig:R_vs_N_aspect_ratios} shows $\mathcal R(N)$ for several representative trap anisotropies. For each fixed anisotropy, $\mathcal R(N)$ approaches the same thermodynamic-limit value
\begin{equation}
\mathcal R_\infty=1-\frac{3\zeta(3)^2}{4\zeta(2)\zeta(4)}\simeq 0.3913,
\end{equation}
consistent with the known three-dimensional asymptotic result~\cite{Kruk23,Idziaszek2005}. This convergence is remarkably slow: deviations from the asymptotic value remain visible even for particle numbers as large as $10^{10}$. 

\begin{figure}
\centering
\includegraphics[width=0.95\linewidth]{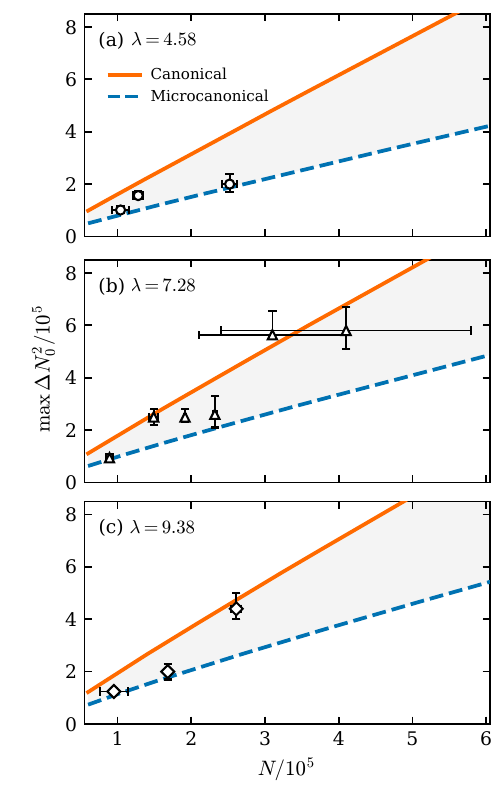}
\caption{%
Comparison with experiment for axially symmetric three-dimensional harmonic traps. Solid and dashed lines show the canonical and microcanonical ideal-gas condensate-number fluctuations maxima, respectively, while symbols denote the experimental measurements of Ref.~\cite{Arlt2021} for the indicated aspect ratios $\lambda=\omega_\rho/\omega_a$. Shading marks the interval between the CE and MCE predictions. Error bars show the reported $68.2\%$ confidence intervals.
}
\label{fig:experiment-comparison}
\end{figure}

To connect these results with experiment, Fig.~\ref{fig:experiment-comparison}
compares the canonical and microcanonical maxima of the condensate-number
variance with the measurements of Ref.~\cite{Arlt2021} for the corresponding
trap aspect ratios. The experimental values lie clearly below the canonical result indicating the microcanonical nature of the fluctuations. Note that residual technical noise in the atom number preparation and detection is always expected to yield a variance of the condensed atom number above the theoretically calculated microcanonical result~\cite{Arlt2021}.

\subsection{Correlation functions in different statistical ensembles}
\label{subsec:correlation-functions-ensembles}

To calculate $g^{(2)}(\mathbf{r}) \equiv g^{(2)}(\mathbf{r}, \mathbf{0}; \mathbf{r}, \mathbf{0})$ in the GCE (see Eq.~\ref{eq:normalized-g}), we follow the approach by Glauber~\cite{NaraschewskiGlauber1999}. This function is then used to obtain the corresponding correlation functions in the CE and MCE. The ensemble dependence is dominated by condensate-number fluctuations, so we separate the condensate mode from the excited modes,
\begin{equation}
\hat\Psi(\mathbf r)
=
u_0(\mathbf r)\hat a_0
+
\delta\hat\Psi(\mathbf r),
\qquad
\delta\hat\Psi(\mathbf r)
=
\sum_{j\geq 1}u_j(\mathbf r)\hat a_j,
\label{eq:field-split-correlation}
\end{equation}
where $u_j(\mathbf r)$ is a single-particle basis function and
$\hat a_j$ annihilates a boson in mode $j$.

We define the grand canonical thermal density and the thermal one-body correlation function:
\begin{equation}
\begin{aligned}
\rho_{\rm T}(\mathbf r)
&=
\sum_{j\geq 1}
|u_j(\mathbf r)|^2
\avggce{n_j},
\\
G_{\rm T}^{(1)}(\mathbf r;\mathbf r')
&=
\sum_{j\geq 1}
u_j^*(\mathbf r)u_j(\mathbf r')
\avggce{n_j}.
\end{aligned}
\label{eq:thermal-one-body-contour}
\end{equation}
\\
For the ideal gas in the grand canonical ensemble there is no correlation between the condensate and the other modes, and the density is (at fixed condensate occupation $n_0=n$) 
\begin{equation}
\rho(\mathbf r,n)
=
n|u_0(\mathbf r)|^2+\rho_{\rm T}(\mathbf r).
\label{eq:conditional-density-contour}
\end{equation}

In the canonical ensemble, any observable $\mathcal O(n,w)$ can be averaged by
inserting it into the contour integral of
Eq.~\eqref{eq:ZNfinal}. Thus,
\begin{equation}
\avgce{\mathcal O}
=
\frac{
\operatorname{Re} I_N[\mathcal O]
}{
\operatorname{Re} I_N[1]
},
\label{eq:canonical-contour-observable-average}
\end{equation}
where
\begin{equation}
I_N[\mathcal O]
=
\frac{1}{2\pi\sqrt{\kappa_2}}
\sum_{n=0}^{N}
\int_{\mathcal I_t^+}
\mathcal O\!\left(w(t),n\right)
e^{\tilde f_{\rm ex}(t,n)}
\,\mathrm dt .
\label{eq:canonical-contour-observable-functional}
\end{equation}
The canonical density and correlation functions are therefore obtained as
\begin{equation}
\begin{aligned}
\rho_N(\mathbf r)
&=
\avgce{\rho(\mathbf r,n)},
\\
G_N^{(p)}
&(\mathbf r_1,\ldots,\mathbf r_p;
\mathbf r'_1,\ldots,\mathbf r'_p)
\\
&=
\avgce{
G_n^{(p)}(\mathbf r_1,\ldots,\mathbf r_p;
\mathbf r'_1,\ldots,\mathbf r'_p)
}.
\end{aligned}
\label{eq:canonical-correlation-contour-average}
\end{equation}
This is the same operation as in
Eq.~\eqref{eq:canonical-moments}, but with the factor $n^m$
replaced by the corresponding observable. The microcanonical ensemble is treated analogously:
\begin{equation}
\begin{aligned}
\avgmce{\mathcal O}
&=
\frac{
\operatorname{Re} I_{N,E}[\mathcal O]
}{
\operatorname{Re} I_{N,E}[1]
},
\\
I_{N,E}[\mathcal O]
&=
\frac{1}{(2\pi)^2\sqrt{\kappa_2\kappa_{02}}}
\sum_{n=0}^{N}
\int_{\mathcal D}
\mathcal O\!\left(w(t),b(s),n\right)
\\
&\qquad\times
e^{\tilde s_{\rm ex}(t,s,n)}
\,\mathrm dt\,\mathrm ds .
\end{aligned}
\label{eq:mce-double-contour-observable}
\end{equation}

\begin{figure*}
\centering
\includegraphics[width=0.95\linewidth]{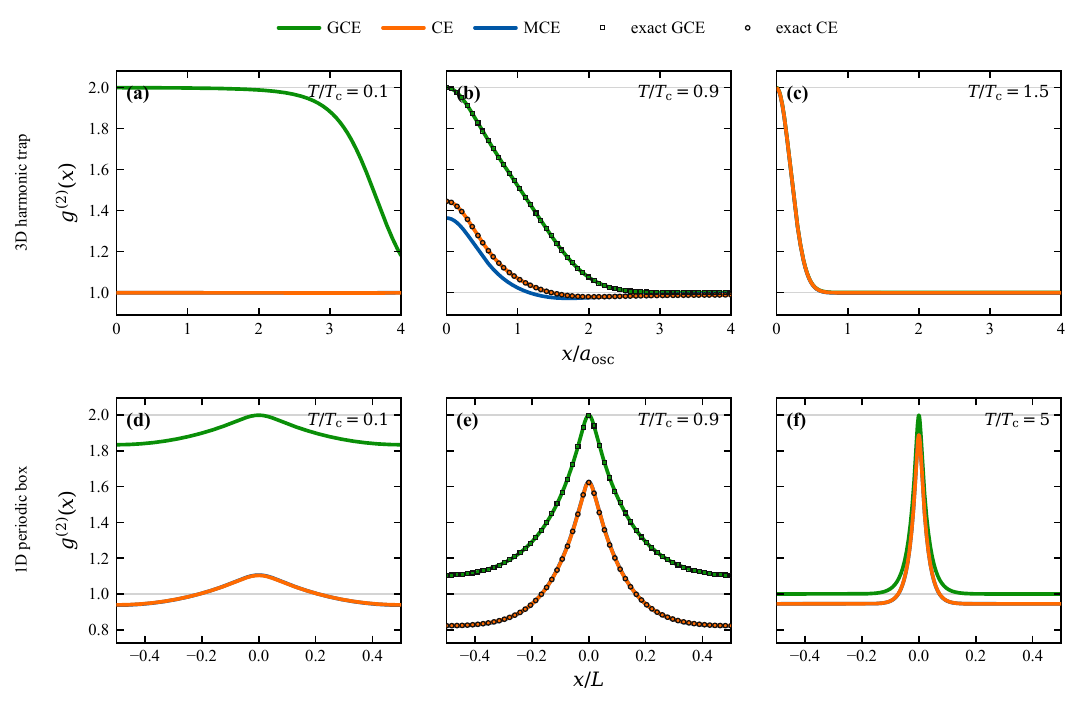}
\caption{
Second-order correlation function $g^{(2)}$ for an ideal Bose gas
with $N=1000$ particles in different temperature regimes.
Top row (a)--(c): $g^{(2)}(x,0)$ for an isotropic
three-dimensional harmonic trap at
$T/T_{\mathrm c}=0.1$, $0.9$, and $1.5$, respectively.
Bottom row (d)--(f): $g^{(2)}(x,0)$ for a one-dimensional periodic box
at $T/T_{\mathrm c}=0.1$, $0.9$, and $5$, respectively.
For the one-dimensional box, $T_{\mathrm c}$ denotes the finite-size
crossover temperature defined by
$\langle n_0\rangle(T_{\mathrm c})=N/2$.
The green, orange, and blue solid curves denote the grand canonical,
canonical, and microcanonical results, respectively.
Open squares and circles in panels (b) and (e) show the independently
evaluated exact grand canonical and canonical results, respectively.
}
\label{fig:g2-temperature-ensemble-comparison}
\end{figure*}

At fixed condensate occupation $n$, the condensate contribution is
evaluated exactly, giving the factor $n(n-1)$. The
excited states form a noninteracting thermal Bose gas, for which
higher-order correlation functions factorize according to Wick's
theorem. We therefore obtain
\begin{align}
G^{(2)}(\mathbf r,\mathbf r')
={}&
n(n-1)
|u_0(\mathbf r)|^2|u_0(\mathbf r')|^2
\nonumber\\
&+
n\Bigl[
|u_0(\mathbf r)|^2\rho_{\rm T}(\mathbf r')
+
|u_0(\mathbf r')|^2\rho_{\rm T}(\mathbf r)
\nonumber\\
&\hspace{2.5em}
+
u_0(\mathbf r)u_0^*(\mathbf r')
G_{\rm T}^{(1)}(\mathbf r;\mathbf r')
\nonumber\\
&\hspace{2.5em}
+
u_0^*(\mathbf r)u_0(\mathbf r')
G_{\rm T}^{(1)}(\mathbf r';\mathbf r)
\Bigr]
\nonumber\\
&+
\rho_{\rm T}(\mathbf r)\rho_{\rm T}(\mathbf r')
+
G_{\rm T}^{(1)}(\mathbf r;\mathbf r')
G_{\rm T}^{(1)}(\mathbf r';\mathbf r).
\label{eq:conditional-G2-contour}
\end{align}
The first term is the condensate-condensate contribution, the second line
contains condensate-thermal direct and exchange terms, and the last line is the
thermal direct-plus-exchange contribution. The canonical and microcanonical second-order functions are then
\begin{equation}
\begin{aligned}
G_N^{(2)}(\mathbf r;\mathbf r')
&=
\frac{
\operatorname{Re} I_N[G^{(2)}(\mathbf r,\mathbf r')]
}{
\operatorname{Re} I_N[1]
},
\\
G_{N,E}^{(2)}(\mathbf r;\mathbf r')
&=
\frac{
\operatorname{Re} I_{N,E}[G^{(2)}(\mathbf r,\mathbf r')]
}{
\operatorname{Re} I_{N,E}[1]
}.
\end{aligned}
\label{eq:CE-MCE-G2-contour-final}
\end{equation}
Finally,
\begin{equation}
g_X^{(2)}(\mathbf r;\mathbf r')
=
\frac{
G_X^{(2)}(\mathbf r;\mathbf r')
}{
\rho_X(\mathbf r)\rho_X(\mathbf r')
},
\qquad
X=N,(N,E).
\label{eq:g2-normalized-contour}
\end{equation}
As an independent benchmark, we also evaluate the canonical $g^{(2)}$ using the exact recurrence relation for $Z_N$ introduced in Eq.~\eqref{eq:canonical-recurrence}, together with exact expressions for the canonical one- and two-body reduced density matrices of an ideal Bose gas~\cite{AkkelinSinyukov2016,TsutsuiKita2016}. The comparison is shown in
Fig.~\ref{fig:g2-temperature-ensemble-comparison}.

\subsection{Time-of-flight correlations}
\label{subsec:tof-correlations}

\begin{figure}
    \centering
    \includegraphics[width=0.95\linewidth]{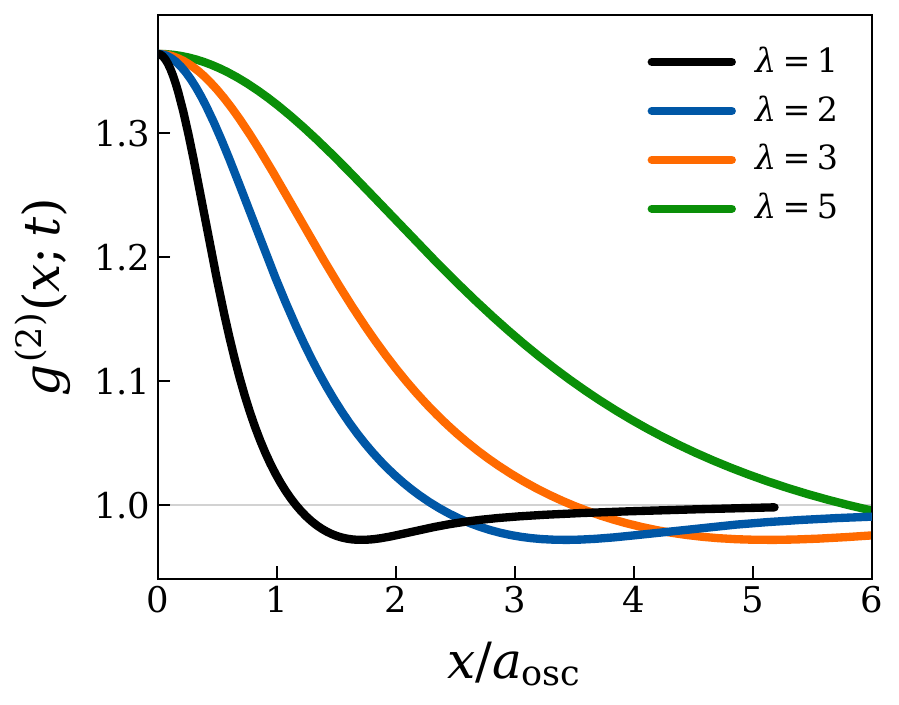}
    \caption{Microcanonical second-order correlation function after free expansion of an ideal Bose gas initially confined in an isotropic three-dimensional harmonic trap, for $N=1000$ and $T/T_{\rm c}=0.9$. The curves correspond to expansion factors $\lambda=1,2,3,$ and $5$, or equivalently to $\omega t=0,\sqrt{3},2\sqrt{2},$ and $2\sqrt{6}$, respectively. The horizontal coordinate is measured in units of the initial oscillator length $a_{\rm osc}$.}
    \label{fig:g2-tof}
\end{figure}

We finally ask how the correlations discussed above are modified after the trap is switched off. 
This is the situation closest to absorption imaging, where the cloud is first allowed to expand and is then detected in real space. 
For a harmonic trap the expansion has a simple scaling form. In the isotropic case used here the relevant length scale is multiplied by
\begin{equation}
    \lambda(t)=\sqrt{1+\omega^2 t^2}.
    \label{eq:tof-lambda}
\end{equation}
Equivalently, the initial oscillator modes are replaced by their freely evolved counterparts,
\begin{equation}
    u_j(\mathbf r,t)
    =
    \lambda^{-3/2}
    \exp\!\left[
        i\,\frac{m\dot{\lambda}}{2\hbar\lambda}\,r^2
    \right]
    e^{-i\varphi_j(t)}
    u_j\!\left(\frac{\mathbf r}{\lambda},0\right),
    \label{eq:tof-modes}
\end{equation}
where the mode-dependent phases $\varphi_j(t)$ cancel from equal-time
one-body correlation functions. The remaining quadratic scaling phase
affects the phase of off-diagonal one-body amplitudes but not their
magnitude.
We then evaluate the same normalized function as in Eq.~\eqref{eq:normalized-g}, using the time-dependent modes in Eq.~\eqref{eq:tof-modes}. 
In particular, we plot $g^{(2)}(x;t)$. The result is shown in Fig.~\ref{fig:g2-tof}. 

The horizontal axis in Fig.~\ref{fig:g2-tof} is given in units of the initial oscillator length $a_{\rm osc}$. During the ballistic expansion, the spatial profile of $g^{(2)}$ broadens by the factor $\lambda$, while its shape is preserved in the scaled coordinate $x/(\lambda a_{\rm osc})$. In particular, the value at the center remains unchanged. The scaling relation holds independently for the GCE, CE, and MCE. Figure~\ref{fig:g2-tof} illustrates this spatial magnification for the microcanonical result; the corresponding CE and GCE profiles undergo the same self-similar transformation.

\subsection{Experimental correlation function measurement}
\label{Experiments}
In experiments with ultracold atoms and Bose-Einstein condensates, the correlation functions up to third order are indeed accessible. The first-order coherence function $g^{(1)}$ can be obtained from interferometric experiments, which typically do not require high spatial resolution of the detection system \cite{Hellweg2003}. Two- and three-body collisional processes are inherently dependent on the correlation between the involved particles and thus give an indirect access to $g^{(2)}$ and $g^{(3)}$ through loss measurements \cite{Kinoshita2005, Haller2011}. A direct measurement of the higher-order correlations, on the other hand, typically requires high resolution and single-atom sensitivity to obtain the correlation function.  %
Here we focus on the direct measurement of $g^{(2)}$ and review the available experimental techniques.%

The general technical challenge is to directly measure a 3-dimensional density distribution as opposed to the typical column density integrated along the line of sight, while maintaining sufficient spatial resolution to resolve the structure of $g^{(2)}$. Figure~\ref{fig:g2-temperature-ensemble-comparison} shows that the required spatial resolution in-situ is a few times smaller than the oscillator length. %
After a time-of-flight expansion, however, the correlation length grows as shown in Fig.~\ref{fig:g2-tof}. A further criterion is the sensitivity to the expected level of density fluctuations. For the following estimation, we assume approximately constant density in the central region of the cloud, where we expect significant correlations. Hence, one needs to resolve density fluctuations on the order of $\delta\rho = \rho(0) \, \sqrt{g^{(2)} - 1}$. %
While it is possible to directly image 3-dimensional atom distribution via tomographic imaging, i.e. imaging selected slices of a cloud \cite{Zwierlein2014, Hong2026}, it has not been employed to measure $g^{(2)}$ and achieving sufficient resolution is challenging. %

Using an inherently two-dimensional system eliminates the problem of determining the density distribution along the line of sight. Combined with a high-resolution imaging system the density-density correlations have been extracted from an in-situ image \cite{Hung2011}. The in-situ detection has been further enhanced by pinning atoms from a bulk cloud to discrete positions with an optical lattice and employing a quantum gas microscope with single-atom sensitivity \cite{Xiang2025}.

Alternatively, a method for mapping out the 3-dimensional atom positions after a time-of-flight has been demonstrated for metastable $^{4}\mathrm{He}^*$ atoms. The atoms fall onto a time- and position-sensitive multi-channel plate detector: their time of arrival contains information about the initial momentum perpendicular to the detector plane, while the detected position maps to the momentum in the plane of the detector. The method has been successfully applied for extracting density-density correlations by counting pairs of atoms hitting the detector at the same position as a function of the difference of their arrival times \cite{Schellekens2005}. The same technique has also been used to additionally extract the third-order correlation function \cite{Hodgman2011}.

\label{subsec:experimental}

\section{Conclusions}
\label{sec:conclusions}

We have studied how the choice of statistical ensemble affects condensate-number fluctuations and spatial correlations in an ideal Bose gas. To access the canonical and microcanonical ensembles at large particle numbers, we introduced a segmented-contour method in which the ground-state contribution is treated exactly, while the excited-state part of the contour is approximated locally. Unlike a single saddle-point expansion, this construction retains the non-Gaussian structure of the condensate statistics. For the one-dimensional harmonic trap, where this non-Gaussianity survives in the thermodynamic limit, the segmented result agrees with exact recurrence calculations while the single-saddle approximation retains a systematic error. In the Gaussian limit the same construction naturally reduces to the usual Gaussian formulas. The microcanonical formulation also reveals a useful separation between particle-number and energy fluctuations. For the ideal gases with power-law spectra, the higher standardized cumulants defined in Eq.~\eqref{eq:standardized-cumulants} and involving the energy vanish in the thermodynamic limit even when the particle-number statistics remain non-Gaussian. The energy projection can therefore be reduced asymptotically to a Gaussian integral while the particle-number contour is still treated by the segmented construction.

These methods allow us to follow ensemble-dependent fluctuations far
beyond the range accessible to exact recurrences. For a three-dimensional
harmonic trap, the ratio $\mathcal R(N)$ between the maximal microcanonical
and canonical condensate-number variances approaches the thermodynamic-limit
value %
$\mathcal R_\infty =
1-3\zeta(3)^2/[4\zeta(2)\zeta(4)]\approx 0.3913$
remarkably slowly. Finite-size corrections remain visible even at $N=10^{10}$, and their magnitude depends strongly on the trap anisotropy, although the asymptotic value itself does not. This slow approach is important when thermodynamic-limit predictions are compared with finite trapped gases.

We have also shown that the choice of ensemble is reflected in the
second-order correlation function $g^{(2)}$. The effect is most pronounced
in the condensed and crossover regimes, where fixing the particle number,
and subsequently the energy, changes the occupation statistics even when
the mean density is very similar. The same contour projection can be
applied to other correlation functions, although here we concentrated on
$g^{(2)}$. These results show that second-order correlations retain
signatures of the statistical constraints imposed on the gas.

To allow for a comparison with experiments that measure $g^{(2)}(x)$ after time-of-flight expansion, we %
argue that the normalized correlations within the cloud profile expand self-similarly: the spatial scale grows with the cloud while the ensemble-dependent structure is preserved. In particular, $g^{(2)}$ obtained in the microcanonical ensemble for a partially condensed cloud shows some bunching $g^{(2)}(0)>1$, which remains visible also at experimentally relevant system sizes after expansion. A natural next step is to extend the present framework to weakly interacting Bose gases and to quantify how interactions and realistic detection effects modify the ensemble dependence of condensate fluctuations and correlation functions.

Phenomenological arguments based on energy-scale comparisons suggest that, in homogeneous systems, BEC fluctuations are strongly suppressed and the canonical and grand canonical ensembles are equivalent. However, most available results focus on solvable 1D systems within the grand canonical ensemble~\cite{Kheruntsyan03,Deuar09,Panfil2014}, and a comprehensive study remains lacking.

\appendix
\section{Details of the segmented-contour method}
\label{app:segmented-method}

Here we give the explicit expansions used in the segmented-contour
method. In the microcanonical case, we also justify the Gaussian approximation
of the energy contour. Throughout the Appendix, we write
\begin{equation}
\Xi_{\rm ex}(w,b)
=
\prod_{\ell\geq1}
\left(1-e^{w-bq_\ell}\right)^{-1},
\qquad
q_\ell=\frac{\varepsilon_\ell}{\delta\varepsilon}.
\label{app:eq:Xi-ex}
\end{equation}

\subsection{Local canonical expansion}
\label{app:subsec:canonical}

On the $j$th segment let
\begin{equation}
\begin{aligned}
t_j&=j\Delta_t,
 \quad
w_j=w_0+i\frac{t_j}{\sqrt{\kappa_2}},
\quad \phi=t-t_j\in[0,\Delta_t].
\end{aligned}
\end{equation}
and define the local derivatives
\begin{equation}
\kappa_r^{(j)}
=
\left.
\partial_w^r\log\Xi_{\rm ex}(w,b_0)
\right|_{w=w_j}.
\label{app:eq:local-number-cumulants}
\end{equation}

For $j>0$, they are generally complex. Since the dependence of
$f_{\rm ex}$ on $n$ is linear, its Taylor expansion can be written as
\begin{equation}
\begin{aligned}
&f_{\rm ex}\!\left(
w_j+i\frac{\phi}{\sqrt{\kappa_2}},n
\right) =Q_j^{(t)}(\phi,n)+R_{t,j}(\phi),
\\[3pt]
&Q_j^{(t)}(\phi,n)=f_{\rm ex}(w_j,n) \, +i\frac{\kappa_1^{(j)}-(N-n)}
{\sqrt{\kappa_2}}\,\phi \, -\frac{\kappa_2^{(j)}}{2\kappa_2}\phi^2,
\\[3pt]
&R_{t,j}(\phi)=
\sum_{r=3}^{\infty}
\frac{i^r\kappa_r^{(j)}}
{r!\,\kappa_2^{r/2}}\phi^r.
\end{aligned}
\label{app:eq:local-CE-expansion}
\end{equation}
In particular, the local quadratic coefficient is
$-\kappa_2^{(j)}/(2\kappa_2)$; replacing
$\kappa_2^{(j)}/\kappa_2$ by unity would be an additional approximation
and is not part of the segmented construction.

The piecewise-quadratic integrand used in
Eqs.~\eqref{eq:ZNfinal} and \eqref{eq:canonical-moments} is therefore
\begin{equation}
e^{\tilde f_{\rm ex}(t,n)}
:=
\sum_{j=0}^{M_t-1}
\mathbf 1_{\mathcal T_j}(t)
\exp\!\left[Q_j^{(t)}(t-t_j,n)\right],
\label{app:eq:fex-piecewise}
\end{equation}
with zero value outside the retained segments.  A concise accuracy
condition is
\begin{equation}
\epsilon_t
:=
\max_{0\leq j<M_t}
\sup_{0\leq \phi\leq\Delta_t}
|R_{t,j}(\phi)|
\ll1.
\label{app:eq:CE-remainder-condition}
\end{equation}
At the central segment its leading term is bounded by
$\gamma_3\Delta_t^3/6$, yielding Eq.~\eqref{eq:conditionsDelta}.

\subsection{Local microcanonical expansion}
\label{app:subsec:microcanonical}

For the second contour introduce
\begin{equation}
\begin{aligned}
s_k&=k\Delta_s,
&
b_k&=b_0+i\frac{s_k}{\sqrt{\kappa_{02}}},
\\
\theta&=s-s_k\in[0,\Delta_s].
\end{aligned}
\end{equation}
and the local mixed derivatives
\begin{equation}
\kappa_{pq}^{(j,k)}
=
(-1)^q
\left.
\partial_w^p\partial_b^q
\log\Xi_{\rm ex}(w,b)
\right|_{(w,b)=(w_j,b_k)}.
\label{app:eq:local-mixed-cumulants}
\end{equation}
With $\phi=t-t_j$ and $\theta=s-s_k$, the second-order polynomial on
$\mathcal T_j\times\mathcal S_k$ is
\begin{align}
Q_{jk}^{(ts)}(\phi,\theta,n)
={}&
s_{\rm ex}(w_j,b_k,n)
+i\frac{\kappa_{10}^{(j,k)}-(N-n)}
{\sqrt{\kappa_2}}\phi
\notag\\
&+i\frac{U-\kappa_{01}^{(j,k)}}
{\sqrt{\kappa_{02}}}\theta
-\frac{\kappa_{20}^{(j,k)}}{2\kappa_2}\phi^2
\notag\\
&+\frac{\kappa_{11}^{(j,k)}}
{\sqrt{\kappa_2\kappa_{02}}}\phi\theta
-\frac{\kappa_{02}^{(j,k)}}{2\kappa_{02}}\theta^2.
\label{app:eq:local-MCE-quadratic}
\end{align}
Thus the explicit approximation entering Eq.~\eqref{eq:OmegaNfinal} is
\begin{align}
e^{\tilde s_{\rm ex}(t,s,n)}
:={}&
\sum_{j=0}^{M_t-1}
\sum_{k=-M_s}^{M_s-1}
\mathbf 1_{\mathcal T_j}(t)
\mathbf 1_{\mathcal S_k}(s)
\notag\\
&\times
\exp\!\left[
Q_{jk}^{(ts)}(t-t_j,s-s_k,n)
\right],
\label{app:eq:sex-piecewise}
\end{align}
and vanishes outside the retained subdomains. Equation
\eqref{app:eq:local-MCE-quadratic} retains both the local curvatures and
the local number--energy covariance $\kappa_{11}^{(j,k)}$.

It remains to control the mixed cubic terms.  Expanding
Eq.~\eqref{app:eq:Xi-ex}, for $p+q=m\geq1$ one obtains at the real saddle point
\begin{equation}
\kappa_{pq}
=
\sum_{\ell\geq1}\sum_{r\geq1}
r^{m-1}q_\ell^q
e^{r(w_0-b_0q_\ell)}.
\label{app:eq:mixed-cumulant-series}
\end{equation}
At a point $(w_j,b_k)$ on the vertical contours, the same
expansion gives
\begin{equation}
\kappa_{pq}^{(j,k)}
=
\sum_{\ell\geq1}\sum_{r\geq1}
r^{m-1}q_\ell^q
e^{r(w_0-b_0q_\ell)}
e^{ir\left(
t_j/\sqrt{\kappa_2}
-q_\ell s_k/\sqrt{\kappa_{02}}
\right)} .
\end{equation}
Hence, by the triangle inequality,
\begin{equation}
\left|\kappa_{pq}^{(j,k)}\right|
\leq
\kappa_{pq}.
\label{app:eq:local-cumulant-bound}
\end{equation}
Since the summands in Eq.~\eqref{app:eq:mixed-cumulant-series} are non-negative, H\"older's inequality with exponents $m/p$ and $m/q$ gives
\begin{equation}
\kappa_{pq}
\leq
\kappa_m^{p/m}\kappa_{0m}^{q/m},
\qquad p+q=m,
\label{app:eq:Holder}
\end{equation}
and hence
\begin{equation}
\gamma_{pq}
\leq
\gamma_m^{p/m}
\gamma_{0m}^{q/m}.
\label{app:eq:Holder-standardized}
\end{equation}
Setting $m=3$ in Eq.~\eqref{app:eq:Holder-standardized} and combining
it with Eq.~\eqref{app:eq:local-cumulant-bound} gives the bound on the
complete cubic contribution used in the main text. On every retained subdomain, the complete cubic contribution therefore obeys
\begin{align}
|R_3(\phi,\theta)|
\leq
\frac{1}{6}
\left(
\gamma_3^{1/3}|\phi|
+\gamma_{03}^{1/3}|\theta|
\right)^3.
\label{app:eq:cubic-MCE-bound}
\end{align}
This justifies Eq.~\eqref{eq:MCE-third-order-bound} and shows that no
independent assumptions on $\gamma_{21}$ and $\gamma_{12}$ are needed.
Together with the analyticity condition
Eq.~\eqref{eq:MCE-analyticity-condition}, it gives the practical
criterion Eq.~\eqref{eq:MCE-cubic-condition}.  As in the canonical case,
the complete Taylor remainder and the omitted contour tails must remain
negligible on the retained domain.

\subsection{Gaussian reduction of the energy contour}
\label{app:subsec:energy-reduction}

Consider a power-law spectrum
$\varepsilon_{\mathbf q}\propto|\mathbf q|^\sigma$ in $d$ independent
quantum-number directions and set
\begin{equation}
\alpha=\frac{d}{\sigma},
\qquad
g(x)\sim Cx^{\alpha-1}.
\end{equation}
In the condensed thermodynamic limit, $w_0\to0$ and $b_0\to0$.
Using Eq.~\eqref{app:eq:mixed-cumulant-series} and replacing the mode sum
by the density of states gives
\begin{equation}
\kappa_{pq}
\asymp
\begin{cases}
b_0^{-(\alpha+q)}, & p<\alpha,\\
b_0^{-(p+q)}\log(b_0^{-1}), & p=\alpha,\\
b_0^{-(p+q)}, & p>\alpha,
\end{cases}
\label{app:eq:power-law-scaling}
\end{equation}
up to positive $b_0$-independent coefficients. It follows that
\begin{equation}
\gamma_m\longrightarrow0\quad(m\geq3)
\quad\Longleftrightarrow\quad
\alpha\geq2,
\label{app:eq:number-Gaussianity}
\end{equation}
where the case $\alpha=2$ converges logarithmically.  More importantly,
for every $\alpha>0$,
\begin{equation}
\gamma_{pq}\longrightarrow0,
\qquad
q\geq1,\qquad p+q\geq3.
\label{app:eq:energy-Gaussianity}
\end{equation}
Thus all standardized higher-order cumulants containing at least one
energy index vanish even when the pure particle-number sector remains
non-Gaussian.

To obtain the single-contour representation, we expand the
exponent to quadratic order in the energy-conjugate variable
$b$, while treating its dependence on $t$ exactly.
We introduce the $t$-dependent energy cumulants
\begin{equation}
\kappa_{0r}(t)
=
(-1)^r
\left.
\partial_b^r \log\Xi_{\rm ex}(w(t),b)
\right|_{b=b_0}.
\end{equation}
Parameterizing the energy contour as
$b(s)=b_0+is/\sqrt{\kappa_{02}}$, we obtain
\begin{equation}
\begin{aligned}
s_{\rm ex}(w(t),b(s),n)
={}& s_{\rm ex}(w(t),b_0,n)\\
&+i\frac{U-\kappa_{01}(t)}
{\sqrt{\kappa_{02}}}\,s\\
&-\frac{\kappa_{02}(t)}
{2\kappa_{02}}\,s^2
+R_E(t,s).
\end{aligned}
\label{app:eq:energy-expansion}
\end{equation}
Equation~\eqref{app:eq:energy-Gaussianity} implies $
R_E(t,s)=o(1) $ on the relevant fluctuation scale. Provided that the energy-contour contribution is localized around
$s=0$, the resulting quadratic approximation can be integrated
analytically over $s\in\mathbb R$, giving
\begin{align}
I_+^{(m)}
\simeq{}&
\frac{1}{(2\pi)^{3/2}\sqrt{\kappa_2}}
\sum_{n=0}^{N}n^m
\int_{\mathcal I_t^+}
\frac{\mathrm dt}{\sqrt{\kappa_{02}(t)}}
\notag\\
&\times
\exp\!\left[
s_{\rm ex}(w(t),b_0,n)
-\frac{[U-\kappa_{01}(t)]^2}
{2\kappa_{02}(t)}
\right].
\label{app:eq:single-contour}
\end{align}
The square-root branch is continued from its positive value at $t=0$.
The remaining particle-number contour is evaluated with the segmented
construction of Sec.~\ref{app:subsec:canonical}.

If the number sector is Gaussian as well, the central quadratic form is
\begin{equation}
-\frac12\left(t^2-2\rho ts+s^2\right),
\qquad
\rho=\frac{\kappa_{11}}
{\sqrt{\kappa_2\kappa_{02}}}.
\end{equation}
Integrating over $s$ leaves
$-\tfrac12(1-\rho^2)t^2$ and hence the conditional variance
\begin{equation}
\kappa_{\rm eff}
=
\kappa_2(1-\rho^2)
=
\kappa_2-\frac{\kappa_{11}^2}{\kappa_{02}},
\label{app:eq:kappa-effective}
\end{equation}
which yields Eq.~\eqref{eq:Pn0_MCE_gaussian}.
For a three-dimensional harmonic trap ($\alpha=3$),
the Gaussian approximation is valid for both the
particle-number and energy contour integrals.
For a three-dimensional box ($\alpha=3/2$) and a
one-dimensional harmonic trap ($\alpha=1$),
segmentation is required only along the
particle-number contour, while the energy contour
can still be treated within the Gaussian approximation.
This distinction follows from the cumulant scaling
of the ideal gas, which relies on the factorization
in Eq.~\eqref{app:eq:Xi-ex}, and does not
necessarily extend to interacting systems.

\bibliographystyle{apsrev4-2}
\bibliography{refs}
\end{document}